\documentclass[unsortedaddress,nofootinbib,11pt]{revtex4}
\usepackage[utf8]{inputenc}
\usepackage{amsmath,empheq,color}
\usepackage{amsfonts}
\usepackage{amsthm}
\usepackage{amssymb}
\usepackage{graphicx}
\usepackage{hyperref}
\usepackage[normalem]{ulem}
\usepackage{epigraph}
\usepackage{mathrsfs}
\usepackage{bbm}
\usepackage{lineno}
\usepackage{scalerel,stackengine}
\usepackage {tikz}
\usetikzlibrary {positioning}
\definecolor {processblue}{cmyk}{0.96,0,0,0}
\usetikzlibrary{calc}
\usepackage{relsize}
\usetikzlibrary{decorations.pathreplacing}
\tikzset{fontscale/.style = {font=\relsize{#1}}}
\usepackage{extarrows}
\usepackage{caption}
\usepackage{subcaption}
\begin{document}

\title{The relation between the symplectic group $Sp(4, \mathbb{R})$ and its Lie algebra: its application in polymer quantum mechanics}

\author{Guillermo Chac\'on-Acosta}
\email{gchacon@cua.uam.mx} 
\affiliation{Departamento de Matem\'aticas Aplicadas y Sistemas, Universidad Aut\'onoma Metropolitana Cuajimalpa, Vasco de Quiroga 4871, Ciudad de M\'exico 05348, MEXICO}

\author{Angel Garcia-Chung}
\email{alechung@xanum.uam.mx} 
\affiliation{Departamento de F\'isica, Universidad Aut\'onoma Metropolitana - Iztapalapa, \\ San Rafael Atlixco 186, Ciudad de M\'exico 09340, M\'exico}
\affiliation{Universidad Panamericana, \\ Tecoyotitla 366. Col. Ex Hacienda Guadalupe Chimalistac, C.P. 01050 Ciudad de M\'exico, M\'exico}

\begin{abstract}
In this paper, we show the relation between $sp(4,\mathbb{R})$, the Lie algebra of the symplectic group, and the elements of $Sp(4,\mathbb{R})$. We use this result to obtain some special cases of symplectic matrices relevant to the study of squeezed states. In this regard, we provide some applications in quantum mechanics and analyze the squeezed polymer states obtained from the polymer representation of the symplectic group. Remarkably, the polymer's dispersions are the same as those obtained for the squeezed states in the usual representation.
\end{abstract}

\maketitle

\tableofcontents

\section{Introduction}

Squeezed states are broadly used in many areas of physics \cite{walls1983squeezed, schnabel2017squeezed, walls2007quantum, adesso2014continuous, braunstein2005quantum}. Of particular interest is the use of these squeezed states in cosmology \cite{grishchuk1990squeezed, polarski1996semiclassicality, lesgourgues1997quantum, kiefer1998quantum, guth1985quantum, martin2016quantum, albrecht1994inflation, gasperini1993quantum, grain2019squeezing}, specifically when arguing for the emergence of semi-classical behavior in the early universe. Loop Quantum Cosmology (LQC) \cite{ashtekar2003mathematical, bojowald2010canonical, bojowald2011quantum, ashtekar2006quantum, ashtekar2006quantum2} is another scenario in which squeezed states are relevant. There, squeezed states for a single-mode show some of the features of the quantum bounce and closely approximate solutions to the classical Einstein equations \cite{taveras2008corrections, mielczarek2012gaussian, gazeau2013quantum, diener2014numerical, diener2014numerical2, corichi2011coherent}. The squeezed states used in LQC are constructed by hand, imposing the Gaussian form of the states to obtain the squeezing nature of the dispersion relations. Moreover, the states describe systems with only one degree of freedom, i.e., single-mode squeezed states \cite{milburn1984multimode, gerry2005introductory}.

In quantum optics, squeezed states can be used to improve the sensitivity of measurement devices beyond the usual quantum noise limits \cite{simon1988gaussian, braunstein2005quantum, walls2007quantum, adesso2014continuous, ma1990multimode, schnabel2017squeezed, pirandola2009correlation, weedbrook2012gaussian}. They are defined by the squeeze operators' action on coherent states, or the vacuum state \cite{walls2007quantum, adesso2014continuous, braunstein2005quantum}. These operators are defined within the Fock representation using the annihilation and creation operators, or in the Wigner representation, using the Wigner functional. A particular squeezed state used in quantum optics is the two-mode squeezed state which plays a prominent role in the study of entanglement for bipartite systems.  Particularly in the limit when the amount of squeezing is infinitely large, the states become EPR-like states \cite{EPRpaper}.

Based on the relevance that squeezed states play in cosmology, LQC, and quantum optics, one might ask whether there is a relation between them and whether it is possible to obtain squeezed states in LQC the same way squeezed states are defined in quantum optics. Recall that the construction used in LQC for the squeezed states is somewhat artificial and does not correspond to any mechanism in the cosmological events. Hence, exploring whether LQC formulation admits an operator similar to the squeeze operator and whose action on some state yields a squeezed state might pave the way to construct such a mechanism in LQC.

To do so, one must consider that in the LQC, the representation of the operators is not weakly continuous, hence the Fock representation is not suitable for the physical description. Instead, the Schr\"odinger representation, which is the scheme inherited from the quantization procedure, seems to be the natural scheme to be considered \cite{ashtekar2003quantum, corichi2007polymer, flores2013propagators, velhinho2007quantum, pawlowski2014separable, Garcia-Chung:2020cag}. Despite the Schr\"odinger representation of the squeeze operator might be obtained using the representation of the infinitesimal squeeze operator via the exponential map \cite{hall2018theory}, in LQC this cannot be done. In addition to the mathematical challenge that this operation requires in standard quantum mechanics, this is not possible in LQC because there is no infinitesimal representation of the squeeze operator. Therefore, in the present work, we will use the representation of the symplectic group $Sp(2n, \mathbb{R})$ in polymer quantum mechanics \cite{Garcia-Chung:2020cag}. It is worth mentioning that Polymer Quantum Mechanics (PQM) can be considered as a ``toy model'' of LQC because they share Hilbert spaces with the same mathematical structures. As a result, the representation of the symplectic group in PQM is mathematically the same as in LQC.

To analyze the squeeze operator corresponding to the bipartite squeezed states in LQC, we will provide the relation between the Lie algebra of the symplectic group $sp(4,\mathbb{R})$ and the Lie group $Sp(4,\mathbb{R})$. As far as the authors' knowledge, this relation has not been reported before. With this result, we show some specific cases and then move to the analysis within polymer quantum mechanics. Also, this relation allows us to describe the single-mode squeeze operator (specifically, the product of two single-mode operators) as a particular case of a symplectic matrix in $Sp(4,\mathbb{R})$.

We will show that the squeezed states derived in this way for LQC share the same features as those used in quantum optics. In particular, the correlations' structure is the same for both the single-mode and the two-mode squeezed states. However, there is no need for a Gaussian-like structure for the initial states upon which the polymer squeeze operators act and such structure is absent in the polymer squeezed states.

This paper is organized as follows: in section (\ref{LAAnalysis}) we calculate the relation between $sp(4,\mathbb{R})$ and $Sp(4,\mathbb{R})$. In section (\ref{Special cases}) we discuss the isomorphism between $sp(4,\mathbb{R})$ and the second-order polynomial operators ${\cal P}(2, \mathbb{R})$ and provide some examples. In Section (\ref{QUANTREP}), we show some of the applications of the results given in section \ref{Special cases}; in particular, we determine the covariance matrix for the squeezed states in standard quantum mechanics. In Section (\ref{LQCosmo}), we analyze the squeeze operators' representation in polymer quantum mechanics and construct the polymer squeezed state. We also calculate the dispersion relations and show that they are equal to those obtained for the standard squeezed states. We give our conclusion in Section  (\ref{Conclusion}).

\section{$Sp(4,\mathbb{R})$ group analysis and  $sp(4,\mathbb{R})$ Lie algebra} \label{LAAnalysis}

In this section we will detail the relation between an arbitrary element of $sp(4, \mathbb{R})$ and its corresponding element in the group $Sp(4, \mathbb{R})$.  This relation is the main result of this section and has not been reported as far as we know. First, let us introduce some preliminary concepts and notation, which we will use throughout the paper, to make the presentation self-contained.

Let us begin by considering the Poisson manifold $(\mathbb{R}^{2n}, \{ , \} )$ with Poisson bracket for the coordinates $q_j$ and momenta $p_j$  ($j=1,2,3, \dots, n$) given by
 \begin{equation}
 \left\{ q_j , q_k \right\} = 0, \qquad  \left\{ p_j , p_k \right\} =0, \qquad  \left\{ q_j , p_k \right\} = \delta_{j k}.\label{PoissonB}
 \end{equation}

These coordinates are collected using the array $\vec{Y}^T = ( q_1, \, p_1 , \, q_2 , \, p_2 , \, \hdots , \, q_n , \, p_n )$ for which the Poisson bracket (\ref{PoissonB}) takes the form
\begin{equation}
\left\{ \vec{Y} , \vec{Y}^T \right\} = \left( \begin{array}{cccc} {\bf J} & {\bf 0} & \cdots & {\bf 0} \\ {\bf 0} & {\bf J}  & \cdots & {\bf 0} \\
\vdots & \vdots & \vdots & \vdots \\ {\bf 0} & {\bf 0} & \cdots & {\bf J} \end{array}\right)  = {\bf 1}_{n \times n} \otimes {\bf J} ,
\end{equation}
\noindent where the ${\bf 0}$ is the $2\times2$ null matrix, ${\bf 1}_{n\times n}$ is the identity matrix and the matrix ${\bf J}$ is given by
\begin{equation}
{\bf J} = \left( \begin{array}{cc} 0 & 1 \\ -1 & 0 \end{array}\right). \label{MatrixJ}
\end{equation}

The group action over the manifold $\mathbb{R}^{2n}$ is
\begin{equation}
Sp(2n, \mathbb{R}) \times \mathbb{R}^{2n} \rightarrow \mathbb{R}^{2n}; \left( {\bf M}, \vec{ Y} \right) \mapsto \vec{ Y}'^T = {\bf M}\; \vec{Y}^T, \label{2Coord}
\end{equation}
\noindent provided that the matrix ${\bf M}$ satisfies the condition
\begin{equation}
\left( {\bf 1}_{n \times n} \otimes {\bf J} \right) = {\bf M} \left( {\bf 1}_{n \times n} \otimes {\bf J} \right) {\bf M} ^T, \label{2SpCond}
\end{equation}

where ${\bf M}^T$ is the transpose matrix. That is, the symplectic group $Sp(2n, \mathbb{R})$ can be defined as the set of $2n \times 2n$ real matrices satisfying (\ref{2SpCond}) and, additionally, its group action on the Poisson manifold $(\mathbb{R}^{2n}, \{, \})$ is given by (\ref{2Coord}). 
Note that a ``coordinatization'' of $({\mathbb{R}^{2n}}, \{, \})$ different from $\vec{Y}$ yields a condition for the symplectic group matrices different to that in (\ref{2SpCond}). To show this, consider now the array  
$\vec{ X}^T = (\vec{q}^{\; T} \; , \; \vec{p}^{\; T})$ where $\vec{q}^{\; T} =(q_1, q_2, \dots, q_n)$ and $\vec{p}^{\; T}=(p_1, p_2, \dots, p_n)$ are the coordinates on the space $\mathbb{R}^{2n}$. The Poisson bracket for this array is given by
\begin{equation}
\left\{ \vec{X} , \vec{X}^T  \right\} = \left( \begin{array}{cc} {\bf 0} & {\bf 1}_{n \times n} \\ - {\bf 1}_{n \times n} & {\bf 0} \end{array}\right) = {\bf J}  \otimes {\bf 1}_{n \times n} .
\end{equation}

The group action is now given by $Sp(2n, \mathbb{R}) \times \mathbb{R}^{2n} \rightarrow \mathbb{R}^{2n}; \left( \widetilde{\bf M}, \vec{X} \right) \mapsto \vec{X}'$ where $\vec{X}'$ is 
\begin{equation}
\vec{ X}'^T = \widetilde{\bf M} \, \vec{ X}^T, \label{1Coord}
\end{equation}
\noindent and the matrix $\widetilde{\bf M}$ satisfies
\begin{equation}
\left( {\bf J}  \otimes {\bf 1}_{n \times n} \right) = \widetilde{\bf M} \left( {\bf J}  \otimes {\bf 1}_{n \times n} \right) \widetilde{\bf M} ^T. \label{1SpCond}
\end{equation}

Hence, both conditions (\ref{2SpCond}) and (\ref{1SpCond}), can be considered as  definitions for the symplectic group in different ``coordinatizations'' of the phase space $\mathbb{R}^{2n}$. Naturally, both group actions $\widetilde{\bf M}$ and ${\bf M}$ are related via the similarity transformation ${\bf \Gamma}(n)$ \cite{adesso2014continuous} as
\begin{equation}
\widetilde{\bf M} = {\bf \Gamma}(n) \; {\bf M} \; {\bf \Gamma}^{-1}(n), \label{RelationbetweenMs}
\end{equation}
\noindent where ${\bf \Gamma}(n)$ is given by
\begin{equation}
\vec{ X}^T = {\bf \Gamma}(n) \, \vec{ Y}^T , \label{CTransformation}
\end{equation}
\noindent and is such that ${\bf \Gamma}^T(n)  = {\bf \Gamma}^{-1}(n)$. Since the present work concerns the case where $n=2$, it is worth showing the explicit form of ${\bf \Gamma}(2)$ which is
\begin{equation}
{\bf \Gamma}(2)=\left( \begin{array}{cccc} 1 & 0 & 0 & 0 \\ 0 & 0 & 1 & 0 \\ 0 & 1 & 0 & 0 \\ 0 & 0 & 0 & 1\end{array}\right).  \label{CTransformation2}
\end{equation}

Having provided the two group actions over the manifold $\mathbb{R}^{2n}$ using different ``coordinatizations'' and their relation for arbitrary $n$, let us now focus on the symplectic group $Sp(4,\mathbb{R})$. According to (\ref{2SpCond}) this group is given by $4\times 4$ real  matrices ${\bf M}$ for which the following condition holds
\begin{equation}
\left( \begin{array}{cc} {\bf J} & 0 \\ 0 & {\bf J} \end{array} \right) = {\bf M} \left( \begin{array}{cc} {\bf J} & 0 \\ 0 & {\bf J} \end{array} \right) {\bf M}^T ,\label{SpCond}
\end{equation}
\noindent The matrix ${\bf M}$ can be written in block form as
\begin{equation}
{\bf M} :=  \left( \begin{array}{cc} {\bf A} & {\bf B} \\ {\bf C} & {\bf D} \end{array}\right), \label{BlockformM}
\end{equation}
\noindent where the $2 \times 2$ block matrices ${\bf A}$, ${\bf B}$, ${\bf C}$ and ${\bf D}$ satisfy the conditions 
\begin{equation}
{\bf J} = {\bf A} {\bf J} {\bf A}^T +  {\bf B} {\bf J} {\bf B}^T =  {\bf C} {\bf J} {\bf C}^T +  {\bf D} {\bf J} {\bf D}^T , \qquad {\bf 0} =  {\bf A} {\bf J} {\bf C}^T +  {\bf B} {\bf J} {\bf D}^T ,
\end{equation}
\noindent which result from (\ref{SpCond}). 

The Lie algebra of $Sp(4,\mathbb{R})$, denoted as $sp(4,\mathbb{R})$, is given by $4\times4$ matrices ${\bf m}$ such that the exponential map \cite{hall2018theory} of the Lie algebra element ${\bf m}$ yields symplectic matrices ${\bf M}$ close to the identity, i.e., 
\begin{equation}
{\bf M} = e^{\bf m} := {\bf 1} + {\bf m} + \frac{1}{2} {\bf m}^2 + \dots + \frac{1}{n!} {\bf m}^n + \dots   \label{MandS}
\end{equation}

 It can be shown that the matrices in $sp(4,\mathbb{R})$ can be written as the product
\begin{equation}
{\bf m} = \left( \begin{array}{cc} {\bf J} & 0 \\ 0 & {\bf J} \end{array} \right) {\bf L},
\end{equation}
\noindent where ${\bf L}$ is a real symmetric matrix written in block form as
\begin{equation}
{\bf L}= \left( \begin{array}{cc} {\bf a} & {\bf b} \\ {\bf b}^T & {\bf c} \end{array}\right), \label{FormulaforL}
\end{equation}
\noindent and where ${\bf b}$ is a $2\times2$ real matrix, whereas ${\bf a}$ and ${\bf c}$ are also real but $2\times2$ symmetric matrices.

If a matrix ${\bf M}$ can be written as in (\ref{MandS}), then its inverse ${\bf M}^{-1}$, its transpose ${\bf M}^T$ and the $n$-power matrix  $({\bf M})^n$, can be written respectively as follows 
\begin{eqnarray}
{\bf M}^{-1} &=& \exp{ \left[ - \left( \begin{array}{cc} {\bf J} & 0 \\ 0 & {\bf J} \end{array} \right) {\bf L} \right]},  \label{InvMandS} \\
{\bf M}^T &=& - \left( \begin{array}{cc} {\bf J} & 0 \\ 0 & {\bf J} \end{array} \right) \, {\bf M}^{-1} \,\left( \begin{array}{cc} {\bf J} & 0 \\ 0 & {\bf J} \end{array} \right) \label{TMatrixS}, \\
({\bf M})^n &=& \exp{ \left[ \left( \begin{array}{cc} {\bf J} & 0 \\ 0 & {\bf J} \end{array} \right) (n {\bf L}) \right]}. \label{PowerMandS}
\end{eqnarray}

Thus, the Lie algebra multiplication in $sp(4, \mathbb{R})$ is given by the matrix commutator $[,]_{m}$. When this multiplication acts on two arbitrary elements ${\bf m}_1 $ and ${\bf m}_2$ gives the element ${\bf m}_3$ defined as
\begin{equation}
{\bf m}_3 := \left[ {\bf m}_1, {\bf m}_2 \right]_m = \left[ \left( \begin{array}{cc} {\bf J} & 0 \\ 0 & {\bf J} \end{array} \right) {\bf L}_1, \left( \begin{array}{cc} {\bf J} & 0 \\ 0 & {\bf J} \end{array} \right) {\bf L}_2 \right] = \left( \begin{array}{cc} {\bf J} & 0 \\ 0 & {\bf J} \end{array} \right) {\bf L}_3,
\end{equation}
\noindent where the matrix ${\bf L}_3$ is also a real symmetric matrix with components of the form
\begin{equation}
{\bf L}_3 = \left( \begin{array}{cc} {\bf a}_1 {\bf J} {\bf a}_2 + {\bf b}_1 {\bf J} {\bf b}^T_2 - {\bf a}_2 {\bf J} {\bf a}_1 - {\bf b}_2 {\bf J} {\bf b}^T_1 & {\bf a}_1 {\bf J} {\bf b}_2 + {\bf b}_1 {\bf J} {\bf c}_2 - {\bf a}_2 {\bf J} {\bf b}_1 - {\bf b}_2 {\bf J} {\bf c}_1 \\  {\bf b}^T_1 {\bf J} {\bf a}_2 + {\bf c}_1 {\bf J} {\bf b}^T_2 - {\bf b}^T_2 {\bf J} {\bf a}_1 - {\bf c}_2 {\bf J} {\bf b}^T_1 & {\bf c}_1 {\bf J} {\bf c}_2 + {\bf b}^T_1 {\bf J} {\bf b}_2 - {\bf c}_2 {\bf J} {\bf c}_1 - {\bf b}^T_2 {\bf J} {\bf b}_1  \end{array}\right), \label{MatrixL3}
\end{equation}
\noindent hence, ${\bf m}_3$ is clearly an element in $sp(4, \mathbb{R})$.

Up to this point, we introduced the main concepts and notations required to derive the relation between $sp(4, \mathbb{R})$ and its corresponding Lie group $Sp(4,\mathbb{R})$. Let us proceed then to obtain the explicit relation between the block matrices ${\bf A}$, ${\bf B}$, ${\bf C}$ and ${\bf D}$ and the Lie algebra element ${\bf L}$. It is worth noting that the following procedure can be applied to higher-order symplectic groups $Sp(2n, \mathbb{R})$ for $n \geq 3$, being this the main reason for its exposition in this section.

Let us collect the even and odd terms of the expansion in (\ref{MandS}) as follows
\begin{eqnarray}
{\bf M} &=& \left[  {\bf 1} + \frac{1}{2!} {\bf m}^2 + \dots + \frac{1}{(2n)!} {\bf m}^{2n} + \dots 	\right] +  {\bf m} \left[ {\bf 1} + \frac{1}{3!}  {\bf m}^2 + \dots + \frac{1}{(2n+1)!} {\bf m}^{2n} + \dots   \right], \label{Expansion} 
\end{eqnarray}
\noindent where ${\bf m}^2$ takes the form
\begin{equation}
 {\bf m}^2 = \left( \begin{array}{cc} - (\det {\bf a} + \det {\bf b}) {\bf 1}_{2\times2} & {\bf J} {\bf d} \\ - {\bf J} {\bf d}^T & - (\det {\bf b} + \det {\bf c}) {\bf 1}_{2\times2} \end{array}\right), \label{SDefinition}
\end{equation}
\noindent and the matrix ${\bf d}$ is defined as
\begin{equation}
{\bf d} = {\bf a} {\bf J} {\bf b} + {\bf b} {\bf J} {\bf c}.
\end{equation}

As can be seen from the expansion  (\ref{Expansion}), to obtain the expression for ${\bf M}$ we need first to determine ${\bf m}^{2n}$. In Appendix \ref{Appendix1} we obtain the expression for ${\bf m}^{2n}$ given in Eq. (\ref{MatrixSDef}). Let us replace this result in the series expansion (\ref{Expansion}), which, after collecting the even and odd terms, gives the following 
\begin{eqnarray}
{\bf A} &=& \alpha^{(e)} + (\alpha^{(o)} - \beta^{(o)} \det{\bf b}) \, {\bf J} \, {\bf a} + \beta^{(o)} {\bf J} \,{\bf b} \, {\bf J} \, {\bf c}\, {\bf J} \, {\bf b}^T, \label{TerminoA}\\
{\bf B} &=& (\gamma^{(o)} - \beta^{(o)} \det{\bf a}) \, {\bf J} \, {\bf b} + \beta^{(e)} ( {\bf J} \,{\bf a} \, {\bf J} \, {\bf b} +  {\bf J} \,{\bf b} \, {\bf J} \, {\bf c} )+  \beta^{(o)}  {\bf J} \,{\bf a} \, {\bf J} \, {\bf b} \, {\bf J} \, {\bf c}, \label{TerminoB} \\
{\bf C} &=& (\alpha^{(o)} - \beta^{(o)} \det{\bf c}) \, {\bf J} \, {\bf b}^T + \beta^{(e)} ( {\bf J} \,{\bf b}^T \, {\bf J} \, {\bf a} +  {\bf J} \,{\bf c} \, {\bf J} \, {\bf b}^T )+  \beta^{(o)}  {\bf J} \,{\bf c} \, {\bf J} \, {\bf b}^T \, {\bf J} \, {\bf a}, \label{TerminoC} \\
{\bf D} &=& \gamma^{(e)} + (\gamma^{(o)} - \beta^{(o)} \det{\bf b}) \, {\bf J} \, {\bf c} + \beta^{(o)} {\bf J} \,{\bf b}^T \, {\bf J} \, {\bf a}\, {\bf J} \, {\bf b}. \label{TerminoD}
\end{eqnarray}
\noindent The coefficients $\alpha^{(e)} $, $\alpha^{(o)} $, $\beta^{(e)} $, $\beta^{(o)} $, $\gamma^{(e)} $ and $\gamma^{(o)} $ were defined in the appendix \ref{Series}.

These expressions for the matrices ${\bf A}$, ${\bf B}$, ${\bf C}$ and ${\bf D}$ link the components of the Lie algebra element ${\bf L}$ with the corresponding symplectic matrix ${\bf M}$ and constitute the main result of this section. Note also the ``non-linear matrix relation'' between the Lie algebra elements and the group elements, particularly the role of the block matrix ${\bf b}$.

A remarkable and direct application of this result is that it allows us to compute the symplectic eigenvalues of the matrix ${\bf M}$. To do so, recall that the characteristic polynomial for a $4 \times 4 $ matrix ${\bf M} $ with $\det({\bf M}) = 1$ can be written in terms of the trace of its first three powers by the expression
\begin{eqnarray}
&& \Lambda^4 - (\mbox{Tr} ({\bf M}))  \Lambda^3 + \frac{\left[ (\mbox{Tr} ({\bf M}))^2 - (\mbox{Tr} ({\bf M}^2)) \right]}{2} \Lambda^2 + \left[ - \frac{(\mbox{Tr} ({\bf M}))^3}{6} + \frac{(\mbox{Tr} ({\bf M})) (\mbox{Tr} ({\bf M}^2))}{2}  + \right. \nonumber \\
&& \left. - \frac{(\mbox{Tr} ({\bf M}^3))}{3}  \right] \Lambda + 1 = 0.  \label{ChEq}
\end{eqnarray}
\noindent where $\Lambda$ are the eigenvalues of the arbitrary matrix ${\bf M}$ and $\mbox{Tr}({\bf M})$ is the trace of the matrix. 

Using the relations (\ref{TerminoA})--(\ref{TerminoD}) we obtain that the trace $\mbox{Tr}({\bf M})$ is given by
\begin{equation}
\mbox{Tr}({\bf M}) = 2 (\alpha^{(e)} + \gamma^{(e)}) = \cosh\left( \sqrt{\lambda_+} \right) + \cosh\left( \sqrt{\lambda_-} \right),
\end{equation}
\noindent where $\alpha^{(e)}$ and $\gamma^{(e)}$ are given in (\ref{alphapar}) and (\ref{gammapar}), respectively, and eigenvalues $\lambda_\pm$ are given in (\ref{eigenvalues}). Moreover, due to the linearity of the trace and the relation (\ref{PowerMandS}), we can verify that $\mbox{Tr}({\bf M}^n)$ is given by
\begin{equation}
\mbox{Tr}({\bf M}^n) =  \cosh\left( n \sqrt{\lambda_+} \right) + \cosh\left(n \sqrt{\lambda_-} \right),
\end{equation}
\noindent from which we obtain the expressions for $\mbox{Tr}({\bf M}^2)$ and $\mbox{Tr}({\bf M}^3)$.

Inserting all these expressions in (\ref{ChEq}) for a symplectic matrix of the form (\ref{MandS}), and when calculating the roots of the polynomial, we obtain that their eigenvalues are given by
\begin{eqnarray}
\Lambda_1 &=& \cosh \left(\sqrt{\lambda_{+}}\right)-\sinh \left(\sqrt{\lambda_{+}}\right) = e^{-\sqrt{ \lambda_{+}}}, \\
\Lambda_2 &=&  \cosh \left(\sqrt{\lambda_{-}}\right)-\sinh \left(\sqrt{\lambda_{-}}\right) = e^{- \sqrt{ \lambda_{-}}} , \\
\Lambda_3 &=&  \sinh \left(\sqrt{\lambda_{-}}\right)+\cosh \left(\sqrt{\lambda_{-}}\right) = e^{\sqrt{ \lambda_{-}}} , \\
\Lambda_4 &=&  \sinh \left(\sqrt{\lambda_{+}}\right)+\cosh \left(\sqrt{\lambda_{+}}\right) = e^{\sqrt{ \lambda_{+}}},
\end{eqnarray}
\noindent Note that if $\lambda_{+} > \lambda_{-}$, then the eigenvalues are ordered as $\Lambda_1 < \Lambda_2 < \Lambda_3 < \Lambda_4 $ which coincides with the results of Williamson's theorem \cite{Arnold, Gosson2006}. 

In \cite{milburn1984multimode, castanos1987matrix} an alternative (and different) formulation for each of the symplectic group generators was provided. Our approach, however, not only reproduces to the same expressions for the generators but also provides a direct relation with the Lie algebra matrix generators ${\bf a}$, ${\bf b}$ and ${\bf c}$ a point that is absent in \cite{milburn1984multimode, castanos1987matrix}. As a result, we can relate a broader range of Lie algebra elements with their corresponding group elements.

Let us now show some of the relevant matrices and examples in the next section in which this result can be applied. 

\section{Quantum relations and examples}  \label{Special cases}

This section provides three examples where the relation between the Lie algebra element and the group element is explicit. However, before proceeding, let us introduce additional concepts and notations (see \cite{hall2018theory, adesso2014continuous} for more details), which will be relevant for the quantum description.

\subsection{Relation between $sp(4,\mathbb{R})$ and ${\cal P}(2, \mathbb{R})$}

Consider the Lie algebra formed by second-order (operator) polynomials, denoted by ${\cal P}(2, \mathbb{R})$. An arbitrary element $\widehat{s}$ is given as
\begin{eqnarray}
\widehat{s} &=& - \frac{i}{\hbar} \left[ \frac{a_{11}}{2}  \widehat{q}^2_1 + \frac{a_{12}}{2}  (\widehat{q}_1 \widehat{p}_1 + \widehat{p}_1 \widehat{q}_1 )  + \frac{a_{22}}{2}  \widehat{p}^2_1 +  b_{11} \widehat{q}_1 \widehat{q}_2 + b_{12} \widehat{q}_1 \widehat{p}_2 + b_{21} \widehat{p}_1 \widehat{q}_2 + b_{22} \widehat{p}_1 \widehat{p}_2 + \right. \nonumber \\
&& \left. + \frac{c_{11}}{2}  \widehat{q}^2_2 + \frac{c_{12}}{2}  (\widehat{q}_2 \widehat{p}_2 + \widehat{p}_2 \widehat{q}_2 )  + \frac{c_{22}}{2}  \widehat{p}^2_2 \right].\label{AQHamiltonian}
\end{eqnarray}
\noindent  Here, $\widehat{q}_j$ and $\widehat{p}_j$, with $j=1,2$, are the position and momenta operators satisfying the canonical commutation relations $[\widehat{q}_j , \widehat{p}_k] = i \hbar \delta_{j ,k}$, and $a_{ij},\; b_{ij}$ and $c_{ij}$ are all real coefficients. The reason for this notation is that formally $i \hbar \widehat{s}$ is a self-adjoint operator to be represented in a Hilbert space ${\cal H}$, hence the exponential map $e^{\widehat{s}}$ gives rise to a unitary operator in ${\cal H}$. Thus, in this sense, this notation smoothes the way to the quantum representation analysis in section \ref{QUANTREP}.

 It is easy to check that $\widehat{s}$ can be written in the following form
\begin{equation}
\widehat{s} = - \frac{i}{2 \hbar} \widehat{ \bf R}^T \left( \begin{array}{cccc} a_{11} & a_{12} & b_{11} & b_{12} \\ a_{12} & a_{22} & b_{21} & b_{22} \\ b_{11} & b_{21} & c_{11} & c_{12} \\ b_{12} & b_{22} & c_{12} & c_{22} \end{array}\right)   \widehat{ \bf R}, \label{MatrixFormH}
\end{equation}
\noindent that is a symmetric matrix, and where 
\begin{equation}
\widehat{ \bf R}^T = \left( \begin{array}{cccc} \widehat{q}_1 & \widehat{p}_1 & \widehat{q}_2 & \widehat{p}_2 \end{array}\right).
\end{equation}

Instead of the matrix commutator, the Lie algebra multiplication in ${\cal P}(2, \mathbb{R})$ is given by the operator commutator $[,]$. Therefore, the Lie algebra multiplication of two elements ${\widehat{s}_1}$ and ${\widehat{s}_2}$ gives a third element $\widehat{s}_3$ of the form
\begin{equation}
\widehat{s}_3 = \left[ \widehat{s}_1, \widehat{s}_2 \right] = - \frac{i}{2\hbar} \widehat{ \bf R}^T {\bf L}_3 \widehat{ \bf R},
\end{equation}
\noindent where the matrix ${\bf L}_3$ is given by (\ref{MatrixL3}). Due to ${\bf L}_3$ is a symmetric matrix, the operator ${\widehat s}_3$ is clearly in ${\cal P}(2, \mathbb{R})$. Naturally, this result provides the isomorphism between $sp(4, \mathbb{R})$ and ${\cal P}(2, \mathbb{R})$, i.e., the map
\begin{equation}
\iota: sp(4, \mathbb{R}) \rightarrow {\cal P}(2, \mathbb{R}); {\bf m} \mapsto \widehat{s} = \iota({\bf m})  = - \frac{i}{2 \hbar} \widehat{ \bf R}^T  \, {\bf L}  \, \widehat{ \bf R}, \label{IsoMorphism}
\end{equation}
\noindent and this map preserves the linear properties of both Lie algebras, i.e., it is a Lie algebra isomorphism.

An implication of this isomorphism is that due to ${\cal P}(2, \mathbb{R})$ is a Lie algebra isomorphic to $sp(4,\mathbb{R})$, then the exponential map of its elements ($\widehat{s} \mapsto e^{\widehat{s}}$) gives a (quantum) unitary operator ($e^{\widehat{s}}$) which can be seen as the (quantum) unitary representation of $Sp(4,\mathbb{R})$ as showed in the following diagram
\begin{equation}
\begin{array}{ccc} \boxed{sp(4, \mathbb{R})} & \xLongleftrightarrow{\iota} & \boxed{{\cal P}(2, \mathbb{R})} \\ \big\downarrow & & \big\downarrow \\  \boxed{Sp(4, \mathbb{R})} & \longrightarrow & \boxed{\widehat{Sp}(4,\mathbb{R})} \end{array}
\end{equation}

 We can expect that if a representation of ${\cal P}(2, \mathbb{R})$ in a Hilbert space ${\cal H}$ is known, then there is also a representation of $\widehat{Sp}(4,\mathbb{R})$ in ${\cal H}$. However, in some scenarios like in polymer quantum mechanics and LQC, it is not possible to obtain the representation of $\widehat{S}$ out of the representation of $\widehat{s}$ in ${\cal H}$. The reason is that some elements of ${\cal P}(2, \mathbb{R})$ cannot be represented in the corresponding Hilbert space. This difficulty can be overcome if we can represent directly the exponential $e^{\widehat{s}}$ instead of its infinitesimal generator $\widehat{s}$. This approach was done for the case of polymer quantum mechanics in \cite{Garcia-Chung:2020cag}. Consequently, due to the operators in (\ref{AQHamiltonian}) can be used to describe the dynamics of many physical systems ranging from two decoupled quantum harmonic oscillators to the bipartite squeeze operators, a polymer representation of these operators is possible, as we will show in section \ref{LQCosmo}. More details about these aspects will be provided in section \ref{LQCosmo}.

 In the next subsection, we show some of the explicit forms of ${\bf M}$.

\subsection{Examples}

\subsubsection{Case ${\bf a}, {\bf c} \neq {\bf 0}$ and ${\bf b}={\bf 0} $. }

Let us consider the Lie algebra element with ${\bf b}={\bf 0}$ and ${\bf a}, {\bf c} \neq 0$, which, according to the expression (\ref{MatrixFormH}), implies that there is no interaction between the subsystems, that is, $\widehat{s}$ is of the form 
\begin{eqnarray}
\widehat{s} &=& - \frac{i}{2 \hbar} \left[ a_{11}  \widehat{q}^2_1 + a_{12}  (\widehat{q}_1 \widehat{p}_1 + \widehat{p}_1 \widehat{q}_1 )  + a_{22}  \widehat{p}^2_1 + c_{11}  \widehat{q}^2_2 + c_{12}  (\widehat{q}_2 \widehat{p}_2 + \widehat{p}_2 \widehat{q}_2 )  + c_{22}  \widehat{p}^2_2 \right] . \label{Case1Hamiltonian}
\end{eqnarray}

 In this case, ${\bf d} = {\bf 0}$ and $\lambda_+ = - \det{\bf a}$ and $\lambda_- = - \det{\bf c}$.  After inserting ${\bf b}=0$ and the expressions for $\lambda_\pm $ in (\ref{alphapar})-(\ref{gammaimpar}) we obtain the following symplectic matrix 
\begin{equation}
{\bf M}_1 = \left( \begin{array}{cc} \cosh\left( \sqrt{- \det{\bf a}}\right) + \frac{ \sinh\left( \sqrt{- \det{\bf a}}\right)}{\sqrt{-\det{\bf a}}}  \, {\bf J} \, {\bf a} & {\bf 0} \\ {\bf 0} & \cosh\left( \sqrt{- \det{\bf c}}\right) + \frac{ \sinh\left( \sqrt{- \det{\bf c}} \right)}{\sqrt{-\det{\bf c}}}  \, {\bf J} \, {\bf c} \end{array}\right). \label{Casebzero}
\end{equation}
\noindent As can be seen, both block matrices in (\ref{Casebzero}) are elements of $Sp(2, \mathbb{R})$ hence, the Lie algebra elements given by the parameters ${\bf a}$ and ${\bf c}$ can be considered as the Lie algebra generators of  $Sp(2, \mathbb{R})\otimes Sp(2, \mathbb{R}) \subset  Sp(4, \mathbb{R})$. Moreover, the matrix ${\bf M}_1$ is diagonal if and only if ${\bf a}$ and ${\bf c}$ are anti-diagonal matrices, i.e., only when there are no squared terms in (\ref{AQHamiltonian}).

An important symplectic matrix of this type is 
\begin{equation}
\left( \begin{array}{cc} {\bf J} & 0 \\ 0 & {\bf J} \end{array} \right) ,
\end{equation}
\noindent which is often used to derive the transpose matrix as in (\ref{TMatrixS}). One can check that this matrix can be obtained from (\ref{Casebzero}) when ${\bf a}=  {\bf c}= \mbox{diag}( \frac{\pi}{2}, \frac{\pi}{2})$, i.e., 
\begin{equation}
\left( \begin{array}{cc} {\bf J} & 0 \\ 0 & {\bf J} \end{array} \right)  = \exp{ \left[ \frac{\pi}{2} \left( \begin{array}{cc} {\bf J} & 0 \\ 0 & {\bf J} \end{array} \right) \right]}.
\end{equation}

\subsubsection{ Case ${\bf c} = {\bf a} = \mbox{diag}(a_{11}, a_{22})$ and ${\bf b} =$diag$(b_{11}, b_{22}) $.}

In this case the operator $\widehat{s}$ is of the form
\begin{eqnarray}
\widehat{s} &=& - \frac{i}{\hbar} \left[  a_{22} \left( \frac{1}{2}  \widehat{p}^2_1 + \frac{1}{2}  \widehat{p}^2_2 \right) + a_{11} \left( \frac{1}{2}  \widehat{q}^2_1 +  \frac{1}{2}  \widehat{q}^2_2  \right)   +  b_{11} \widehat{q}_1 \widehat{q}_2  + b_{22} \widehat{p}_1 \widehat{p}_2  \right] ,\label{Case2AQHamiltonian}
\end{eqnarray}
\noindent i.e.,  the sub-systems interact via the matrix ${\bf b}$ but only with couplings between coordinates $\widehat{q}_1 \widehat{q}_2$ and momenta operators $\widehat{p}_1 \widehat{p}_2$. According to (\ref{eigenvalues}), the expression for $\lambda_\pm$ for this case is
\begin{equation}
\lambda_{\pm } = - ( a_{11} \, a_{22} + b_{11} \, b_{22} )  \pm (a_{11} \, b_{22} + a_{22} \, b_{11}) = - (a_{11} \mp b_{11}) (a_{22} \mp b_{22}).
\end{equation}
Note that when $b_{11} = \pm a_{11}$ or $b_{22} = \pm a_{22}$ the eigenvalues are null and two particular systems emerge with their operators given by
\begin{eqnarray}
b_{11} = \pm a_{11} \rightarrow \widehat{s} &=& - \frac{i}{\hbar} \left[  a_{22} \left( \frac{1}{2}  \widehat{p}^2_1 + \frac{1}{2}  \widehat{p}^2_2 \right) + \frac{a_{11}}{2} \left(  \widehat{q}_1 \pm   \widehat{q}_2  \right)^2   + b_{22} \widehat{p}_1 \widehat{p}_2  \right] ,\\
b_{22} = \pm a_{22} \rightarrow \widehat{s} &=& - \frac{i}{\hbar} \left[  \frac{a_{22}}{2} \left(   \widehat{p}_1 +  \widehat{p}_2 \right)^2 + a_{11} \left( \frac{1}{2}  \widehat{q}^2_1 +  \frac{1}{2}  \widehat{q}^2_2  \right)   +  b_{11} \widehat{q}_1 \widehat{q}_2    \right].
\end{eqnarray}
\noindent Both systems represent two interacting harmonic oscillators with a coupling term in the momenta and the coordinates, respectively.

The symplectic matrix, denoted in this case as ${\bf M}_2$, is given by

{\small \begin{equation}
{\bf M}_2 = \left(
\begin{array}{cccc}
 \frac{\cosh \left(\sqrt{ \lambda_{-} }\right) + \cosh \left(\sqrt{\lambda_{+} }\right)}{2}  & \frac{ (a_{22} + b_{22}) S_{-} + (a_{22} - b_{22} ) S_{+}}{2}  & \frac{\cosh \left(\sqrt{ \lambda_{-}}\right) - \cosh \left(\sqrt{ \lambda_{+}}\right)}{2}  & \frac{(a_{22}+b_{22}) S_{-} + (b_{22}-a_{22}) S_{+}}{2 }   \\
 \frac{(b_{11}-a_{11}) S_{+} - (a_{11}+b_{11}) S_{-} }{2}    & \frac{\cosh \left(\sqrt{\lambda_{-}}\right) + \cosh \left(\sqrt{\lambda_{+}}\right) }{2}  & \frac{(a_{11}-b_{11}) S_{+} - (a_{11}+b_{11}) S_{-}  }{2}  & \frac{\cosh \left(\sqrt{\lambda_{-}}\right)-\cosh \left(\sqrt{\lambda_+}\right)}{2}  \\
 \frac{\cosh \left(\sqrt{ \lambda_{-}}\right) - \cosh \left(\sqrt{ \lambda_{+}}\right)}{2}  & \frac{(a_{22}+b_{22}) S_{-} + (b_{22}-a_{22}) S_{+}}{2}    & \frac{ \cosh \left(\sqrt{\lambda_{-}}\right)+\cosh \left(\sqrt{\lambda_+}\right)}{2} & \frac{(a_{22}+b_{22}) S_{-} + (a_{22}-b_{22}) S_{+}}{2 }  \\
 \frac{(a_{11}-b_{11}) S_{+} - (a_{11}+b_{11}) S_{-} }{2}  & \frac{ \cosh \left(\sqrt{ \lambda_{-}}\right) - \cosh \left(\sqrt{ \lambda_+}\right) }{2}  & \frac{(b_{11}-a_{11}) S_{+}  - (a_{11}+b_{11}) S_{-}}{2 }  & \frac{\cosh \left(\sqrt{\lambda_{-}}\right)+\cosh \left(\sqrt{\lambda_+}\right)}{2}  \\
\end{array}
\right), \label{SecondExample}
\end{equation}
}
\noindent where we introduce the parameters $S_{\pm}$ as
\begin{equation}
S_{\pm} := \frac{\sinh\left( \sqrt{\lambda_{\pm}} \right)}{ \sqrt{\lambda_{\pm}}}.
\end{equation}

\subsubsection{Case ${\bf a} = {\bf c} = {\bf 0}$ and ${\bf b} \neq {\bf 0}$. } \label{thirdPartCase}
In this case, the operator $\widehat{s}$ is of the form
\begin{eqnarray}
\widehat{s} &=& - \frac{i}{\hbar} \left[   b_{11} \widehat{q}_1 \widehat{q}_2 + b_{12} \widehat{q}_1 \widehat{p}_2 + b_{21} \widehat{p}_1 \widehat{q}_2 + b_{22} \widehat{p}_1 \widehat{p}_2 \right] , \label{Case3AQHamiltonian}
\end{eqnarray}
\noindent and this system corresponds, as we will see in the next section, to the general case of the squeeze operator for a bi-partite system \cite{adesso2014continuous}.

Note that in this case, not only the matrices ${\bf a} = {\bf c} $ are null, but also the matrix ${\bf d}$, which implies that $\lambda_+ = \lambda_- = - \det{\bf b}$. Once we replace these expressions in (\ref{alphapar})-(\ref{gammaimpar}) the symplectic matrix takes the form
\begin{equation}
{\bf M}_3 = \left( \begin{array}{cc} \cosh\sqrt{- \det{\bf b}}  & \frac{\sinh\sqrt{- \det{\bf b}}}{\sqrt{- \det{\bf b}}}   \, {\bf J} \, {\bf b} \\  \frac{\sinh\sqrt{- \det{\bf b}}}{\sqrt{- \det{\bf b}}}   \, {\bf J} \, {\bf b}^T & \cosh\sqrt{- \det{\bf b}}  \end{array}\right),\label{Caseaandczero}
\end{equation}
\noindent where the block matrices ${\bf A}$ and ${\bf D}$ are diagonal matrices. Clearly, when $\det{\bf b} < 0$ the coefficients of ${\bf M}_3$ will be given by hyperbolic functions. In case $\det{\bf b} > 0$ the coefficients are described by trigonometric functions instead.

After giving some examples of symplectic matrices obtained through the exponential map, we are ready to show some of their quantum mechanics applications, both in the standard representation and the so-called polymer or loop representation.

\section{Quantum representation and its applications} \label{QUANTREP}

The unitary representation of the group $Sp(2n, \mathbb{R})$ was given by Moshinsky and Quesne in \cite{moshinsky1971linear}. A review and a historical analysis can be found in \cite{torre2005linear, wolf2016development}. However, to be self-contained, we will show the main aspects of this group's quantum representation in standard quantum mechanics in the next subsection.

\subsection{Schr\"odinger representation of $Sp(2n, \mathbb{R})$}

The symplectic group is a non-compact group which implies an infinite-dimensional Hilbert space for its unitary representation. Consider the Hilbert space ${\cal H} = L^2(\mathbb{R}^n, d\vec{x})$ where $d\vec{x}$ is the standard Lebesgue measure. The unitary representation of $Sp(2n,\mathbb{R})$ is the map 
\begin{equation}
\widehat{C}: Sp(2n,\mathbb{R}) \rightarrow {\cal U}({\cal H}); \; \widetilde{\bf M} \mapsto \widehat{C}_{\widetilde{\bf M}},
\end{equation}
\noindent where $\widehat{C}_{\widetilde{\bf M}}$ is a unitary operator over ${\cal H}$, i.e., formally $\widehat{C}^\dagger_{\widetilde{\bf M}} = \widehat{C}^{-1}_{\widetilde{\bf M}}$. 
Note that the group action considered in this map is $\widetilde{\bf M}$ instead of ${\bf M}$, i.e., we used the ``coordinatization'' given by $\vec{X}$ introduced in section \ref{LAAnalysis}. Hence, in order to obtain a quantum (unitary) representation of a given symplectic matrix ${\bf M}$ we first have to transform it into the other group action $\widetilde{\bf M}$ using Eq. (\ref{RelationbetweenMs}) with the corresponding matrix ${\bf \Gamma}(n)$ given by (\ref{CTransformation}) or ${\bf \Gamma}(2)$ for $Sp(4,\mathbb{R})$ given in (\ref{CTransformation2}). 

The map $\widehat{C}$ is given by the integral operator
\begin{equation}
\widehat{C}_{\widetilde{\bf M}} \Psi(\vec{x}) = \int d\vec{x}' C_{\widetilde{\bf M}}(\vec{x} , \vec{x}') \Psi(\vec{x}'), \qquad \quad \Psi(\vec{x}) \in {\cal H}, \label{UnitaryRep}
\end{equation}
\noindent and the kernel $C_{\widetilde{\bf M}}(\vec{x} , \vec{x}')$ of this integral is 
\begin{equation}
C_{\widetilde{\bf M}}(\vec{x}, \vec{x}') = \frac{e^{ \frac{i}{2 \hbar} \left[ \vec{x}^T {\widetilde{\bf D}} {\widetilde{\bf B}}^{-1} \vec{x} - 2 \vec{x}'^T {\widetilde{\bf B}}^{-1} \vec{x} + \vec{x}'^T {\widetilde{\bf B}}^{-1} {\widetilde{\bf A}} \vec{x}'\right]}}{\sqrt{ (2 \pi i \hbar)^n \det {\widetilde{\bf B}}}} . \label{Kernel}
\end{equation}

According to \cite{moshinsky1971linear}, this representation results from imposing two conditions on the operators $\widehat{C}_{\widetilde{\bf M}}$. The first one is given by 
\begin{equation}
\widehat{C}_{\widetilde{\bf M}}  \left(
\begin{array}{c} 
\vec{\widehat{q}}^T   \\
\vec{\widehat{p}}^T
\end{array}
\right) \widehat{C}^{-1}_{\widetilde{\bf M}} = \widetilde{\bf M}^{-1} \left(
\begin{array}{c} 
\vec{\widehat{q}}^T   \\
\vec{\widehat{p}}^T
\end{array} 
\right), \label{MochisnkyCondition}
\end{equation}
\noindent and relates the symplectic group elements $\widetilde{\bf M}$ with the operators $\widehat{C}_{\widetilde{\bf M}}$. 
\noindent Here, $\vec{\widehat{q}} := (\widehat{q}_1, \widehat{q}_2, \dots, \widehat{q}_n)$ and $\vec{\widehat{p}} := (\widehat{p}_1, \widehat{p}_2, \dots, \widehat{p}_n)$ are the coordinate and momenta operators associated to the Heisenberg Lie algebra of the system. The second condition is that
\begin{equation}
\widehat{C}_{\widetilde{\bf M}} \cdot \left( \widehat{C}_{\widetilde{\bf M}} \right)^{\dagger} = \widehat{1}, \label{UnitaryCond}
\end{equation}
\noindent where $\widehat{1}$ is the identity operator and this results in the unitarity of $\widehat{C}_{\widetilde{\bf M}}$.

 The factor $\det {\bf B}$ in (\ref{Kernel}) gives rise to a well define operator even in the case where the matrix ${\bf B}$ is singular (for more details see \cite{moshinsky1971linear, wolf2016development}). Finally, it is worth mentioning that this representation (\ref{UnitaryRep}) is valid for the entire symplectic group and not just for those elements close to the group identity.

Since the fundamental operators are unbounded the condition (\ref{MochisnkyCondition}) only holds in a subspace given by the domain of the operators $\widehat{q}_j$ and $\widehat{p}_j$ in ${\cal H}$. To obtain a condition valid in the full Hilbert space, we are forced to introduce the exponentiated version of $\widehat{q}_j$ and $\widehat{p}_j$, that is to say, the Weyl algebra. Briefly, the Weyl algebra is a $C^*$-unital algebra whose generators, denoted by $\widehat{W}(\vec{a}, \vec{b})$, are related with $\widehat{q}_j$ and $\widehat{p}_j$ with the following relation
\begin{equation}
\widehat{W}(\vec{a}, \vec{b}) :=  e^{  \frac{i}{\hbar} \left(  \vec{a} \; \vec{\widehat{q}}^{\, T} + \vec{b} \; \vec{\widehat{p}}^{\; T} \right) }, \label{WeylAlgGen}
\end{equation}
\noindent and such that the real arrays $\vec{a} = (a_1, a_2, \dots, a_n)$ and $\vec{b} = (b_1, b_2, \dots, b_n)$, which have dimensions $[a_j] = \mbox{momentum}$ and $[b_j] = \mbox{position}$, label the Weyl algebra generators.

 The standard Schr\"odinger representation of $\widehat{q}_j$ and $\widehat{p}_j$ is now used to obtain a representation for the generators $\widehat{W}(\vec{a}, \vec{b})$ in ${\cal H}$ given by
\begin{equation}
\widehat{W}(\vec{a}, \vec{b}) \Psi(\vec{x}) = e^{ \frac{i}{2 \hbar} \vec{a} \; \vec{b}^{\, T} } e^{ \frac{i}{\hbar} \vec{a} \; \vec{x}^{\,T} } \Psi(\vec{x} + \vec{b}),\label{PCRep}
\end{equation}
\noindent and such that the canonical commutation relations give rise to the Weyl algebra mutiplication
\begin{equation}
\widehat{W}(\vec{a}_1, \vec{b}_1) \widehat{W}(\vec{a}_2, \vec{b}_2) = e^{- \frac{i}{2 \hbar} \left( \vec{a}_1 \;  \vec{b}^T_2 - \vec{b}_1 \; \vec{a}^T_2 \right)} \widehat{W}(\vec{a}_1 + \vec{a}_2, \vec{b}_1 + \vec{b}_2).
\end{equation}

Combining (\ref{MochisnkyCondition}) and (\ref{WeylAlgGen}) to obtain the exponentiated version of (\ref{MochisnkyCondition}) yields
\begin{equation}
\widehat{C}_{\widetilde{\bf M}} \, \widehat{W}(\vec{a}, \vec{b}) \, (\widehat{C}_{\widetilde{\bf M}})^{-1} = \widehat{W}(\vec{a} \; \widetilde{\bf D}^T - \vec{b} \; \widetilde{\bf C}^T , - \vec{a} \; \widetilde{\bf B}^T + \vec{b} \; \widetilde{\bf A}^T),
\end{equation}
\noindent where $\widetilde{\bf A}$, $\widetilde{\bf B}$, $\widetilde{\bf C}$ and $\widetilde{\bf D}$ are the block matrices in $\widetilde{\bf M}$. This relation allows us to obtain a representation of the symplectic group in the Hilbert space used in polymer quantum mechanics and in loop quantum cosmology \cite{Garcia-Chung:2020cag}. 

We are now ready to show, in the next subsections, some of the applications of the representation of $Sp(4,\mathbb{R})$ given by (\ref{UnitaryRep}) and (\ref{Kernel}).

\subsection{Schr\"odinger representation of the squeeze operator for a bi-partite system.}

The squeeze operator $\widehat{S}(\zeta)$ for a bi-partite system is given by the exponential map
\begin{equation}
\widehat{S}(\zeta) = e^{\widehat{s}_\zeta}, \label{SqOperator}
\end{equation}
\noindent where the operator $\widehat{s}_\zeta$, is given by
\begin{equation}
\widehat{s}_\zeta :=  { \frac{1}{2} \left( \zeta^* \widehat{a}_1 \widehat{a}_2 - \zeta \widehat{a}^\dagger_1 \widehat{a}^\dagger_2 \right) }. \label{SqGenOperator}
\end{equation}

Here, $\widehat{a}_1$ and $\widehat{a}_2$ are the annihilation operators for each of the sub-systems, say, 1 and 2, of the bi-partite system, $\widehat{a}^\dagger_1$ and $\widehat{a}^\dagger_2$ are their adjoint operators respectively and $\zeta$ is a complex number labelling the amount of squeezing. The operator $\widehat{S}(\zeta) $, when acting on the vacuum state of the bi-partite quantum harmonic oscillators, gives a family of squeezed states labelled by $\zeta$. 

The operators in (\ref{SqGenOperator}) are in the Fock representation, hence, let us transform (\ref{SqGenOperator}) to the Schr\"odinger representation described with operators $\widehat{q}_1$, $\widehat{q}_2$, $\widehat{p}_1$ and $\widehat{p}_2$. The relation between these representations is given by
\begin{eqnarray}
\widehat{a}_j =  \frac{1}{\sqrt{2}} \frac{\widehat{q}_j}{l_j} + \frac{i  }{\sqrt{2}}  \frac{ l_j \widehat{p}_j}{\hbar} , \qquad \widehat{a}^\dagger_j =  \frac{1}{\sqrt{2}} \frac{\widehat{q}_j}{l_j} - \frac{i  }{\sqrt{2}}  \frac{ l_j \widehat{p}_j}{\hbar} , \label{AniCreOps}
\end{eqnarray}
\noindent for $j=1,2$ and $l_j := \sqrt{\frac{\hbar}{m_j \omega_j}}$ where $m_j$ and $\omega_j$ stand for the masses and the frequencies of the oscillators. Inserting these expressions for $\widehat{a}_j$ and $\widehat{a}^\dagger_j$ in (\ref{SqGenOperator}) the operator $\widehat{s}_{\zeta}$ takes the following form
\begin{eqnarray}
\widehat{s}_{\zeta} &=&  \frac{1}{2i \hbar} \left[  \frac{\hbar \, \zeta_y}{l_1 l_2} \; \widehat{q}_1 \widehat{q}_2 -  \frac{l_2 \, \zeta_x}{l_1} \; \widehat{q}_1 \widehat{p}_2 -  \frac{l_1\, \zeta_x}{l_2} \; \widehat{p}_1 \widehat{q}_2 -  \frac{l_1 l_2 \, \zeta_y}{\hbar} \widehat{p}_1 \widehat{p}_2  \right], \label{SqGen}
\end{eqnarray}
\noindent where $\zeta_x$ and $\zeta_y$ are the real and imaginary parts of $\zeta$. 

We now rewrite this operator in the form
\begin{equation}
\widehat{s}_{\zeta} = - \frac{i}{4\hbar} (\vec{\widehat{R}}^T_1, \vec{\widehat{R}}^T_2) \left( \begin{array}{cc} 0 & {\bf b} \\ {\bf b}^T & 0 \end{array}\right) \left( \begin{array}{c} \vec{\widehat{R}}_1 \\ \vec{\widehat{R}}_2 \end{array}\right), \label{SqueezedGen}
\end{equation}
\noindent where the matrix ${\bf b}$ is the following
\begin{equation}
{\bf b} = \left( \begin{array}{cc}  \frac{\hbar \, \zeta_y}{l_1 l_2} & - \frac{l_2 \,  \zeta_x}{l_1}  \\ -  \frac{l_1\, \zeta_x}{l_2}  & -  \frac{l_1 l_2 \, \zeta_y}{\hbar}  \end{array}\right). \label{bMatrix}
\end{equation}

Using the isomorphism $\iota^{-1}$ defined in (\ref{IsoMorphism}) we obtain that the corresponding Lie algebra element ${\bf m}_\zeta = \iota^{-1}(\widehat{s}_{\zeta})$ is given by
\begin{equation}
{\bf m}_\zeta =  \left( \begin{array}{cc} {\bf J} & 0 \\ 0 & {\bf J} \end{array} \right)\left( \begin{array}{cc} 0 & {\bf b} \\ {\bf b}^T & 0 \end{array}\right).
\end{equation}
\noindent Note that the Lie algebra matrix ${\bf m}_\zeta$ isomorphic to the squeeze operator $\widehat{s}_\zeta$, is of the type given in the third case (\ref{thirdPartCase}).

To obtain the symplectic matrix associated to this Lie algebra element, we insert ${\bf m}_\zeta$ and its expressions for ${\bf a}$, ${\bf b}$ and ${\bf c}$ in (\ref{Caseaandczero}). This results in the following symplectic matrix
\begin{equation}
{\bf M}_{\bf s}(r, \phi) = \left(
\begin{array}{cccc}
 \cosh (r) & 0 & -\sinh (r) \cos (\phi ) \frac{l_1}{l_2}& - \sinh (r) \sin (\phi ) \frac{l_1 l_2}{\hbar}\\
 0 & \cosh (r) & -\sinh (r) \sin (\phi ) \frac{\hbar}{l_1 l_2} & \sinh (r) \cos (\phi ) \frac{l_2}{l_1} \\
 -\sinh (r) \cos (\phi ) \frac{l_2}{l_1} & - \sinh (r) \sin (\phi ) \frac{l_1 l_2}{\hbar} & \cosh(r) & 0 \\
-\sinh (r) \sin (\phi ) \frac{\hbar}{l_1 l_2} & \sinh (r) \cos (\phi ) \frac{l_1}{l_2} & 0 & \cosh (r) \\
\end{array} 
\right), \label{FinalFormMS}
\end{equation}
\noindent  where $r$ and $\phi$ are defined as $\zeta = r e^{i \phi}$. Matrix ${\bf M}_{\bf s}(r, \phi)$ can be considered as the classical symplectic transformation such that when represented in $L^2(\mathbb{R}^2, d^2 \vec{x})$, gives rise to the quantum operator $\widehat{S}(\zeta)$. Naturally, this means also that the unitary representation of $\widehat{S}(\zeta)$ in the Schr\"odinger representation is given by $\widehat{C}_{ {\bf M}_{\bf s} }$, i.e., $\widehat{C}_{ {\bf M}_{\bf s} } = e^{\widehat{s}_\zeta}$.

 It is worth to mention that although the expression (\ref{FinalFormMS}) depends on the proper lengths $l_1$ and $l_2$, the matrix ${\bf M}_{\bf s}(r, \phi)$ is $\hbar-$independent, i.e., it is entirely a classical object. Also, matrix ${\bf M}_{\bf s}$ produces classical squeezing but of course, adapted to the classical phase space, which in this case is $(\mathbb{R}^4, \{, \})$. To illustrate the squeezing and the rotation properties of the matrix ${\bf M}_{\bf s}$ as a canonical transformation for different values of $r$ and $\phi$ we consider its action on a circular trajectory $(q_1(t), p_1(t), q_2(t), p_2(t))$ where, $q_j(t) = \cos(t) \; q_j + \sin(t) \; p_j$ and $p_j(t) = - \sin(t) \; q_j + \cos(t) \; p_j$, for $j=1,2$. 
\begin{figure}[h!]
     \centering
     \begin{subfigure}[b]{0.48\textwidth}
         \centering
         \includegraphics[width=0.4\textwidth]{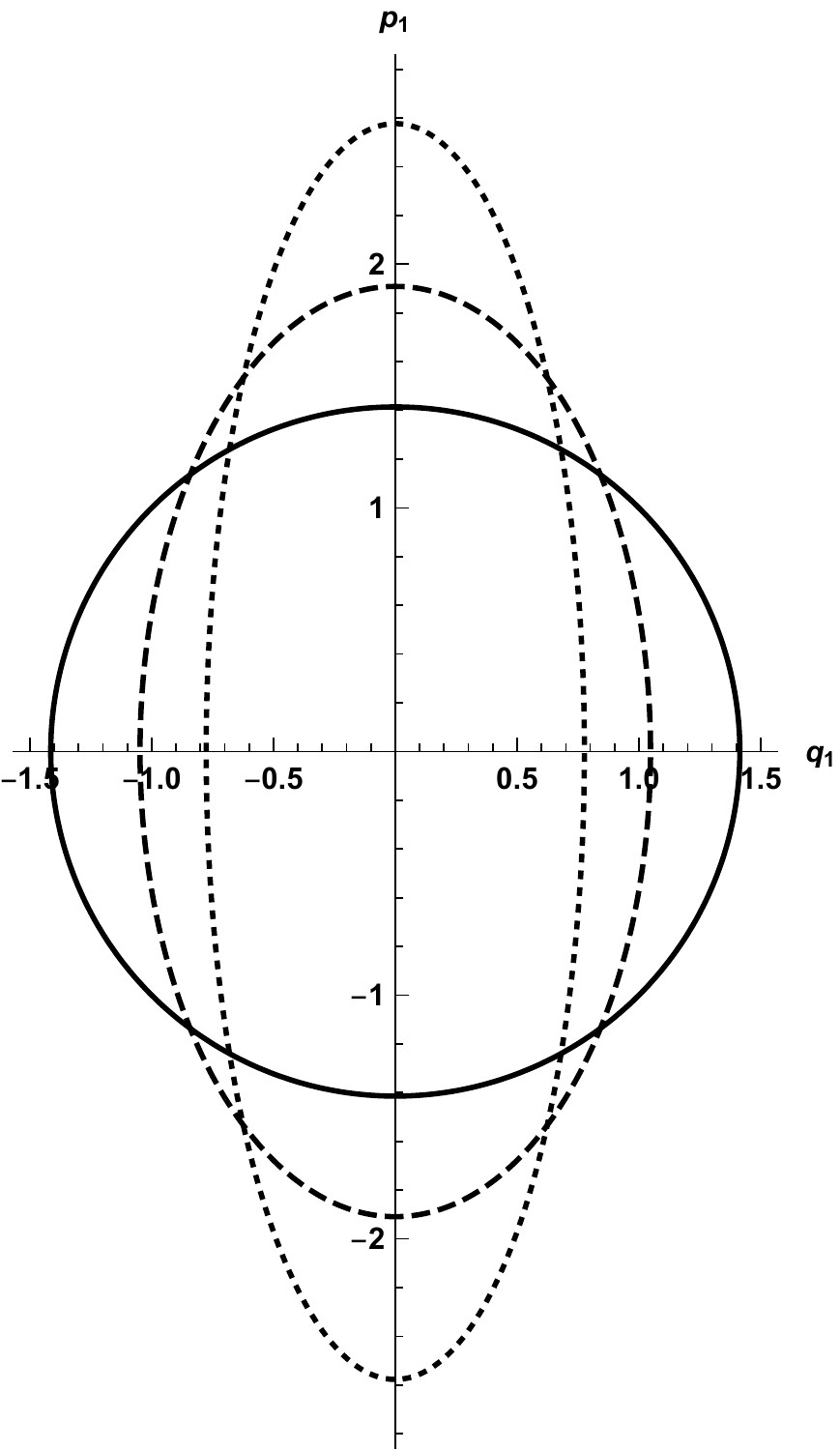}
         \caption{{\footnotesize{Squeezing a circular trajectory.}}}
         \label{Squeezing}
     \end{subfigure}
     \hspace{0.5cm}
     \begin{subfigure}[b]{0.47\textwidth}
         \centering
         \includegraphics[width=0.65\textwidth]{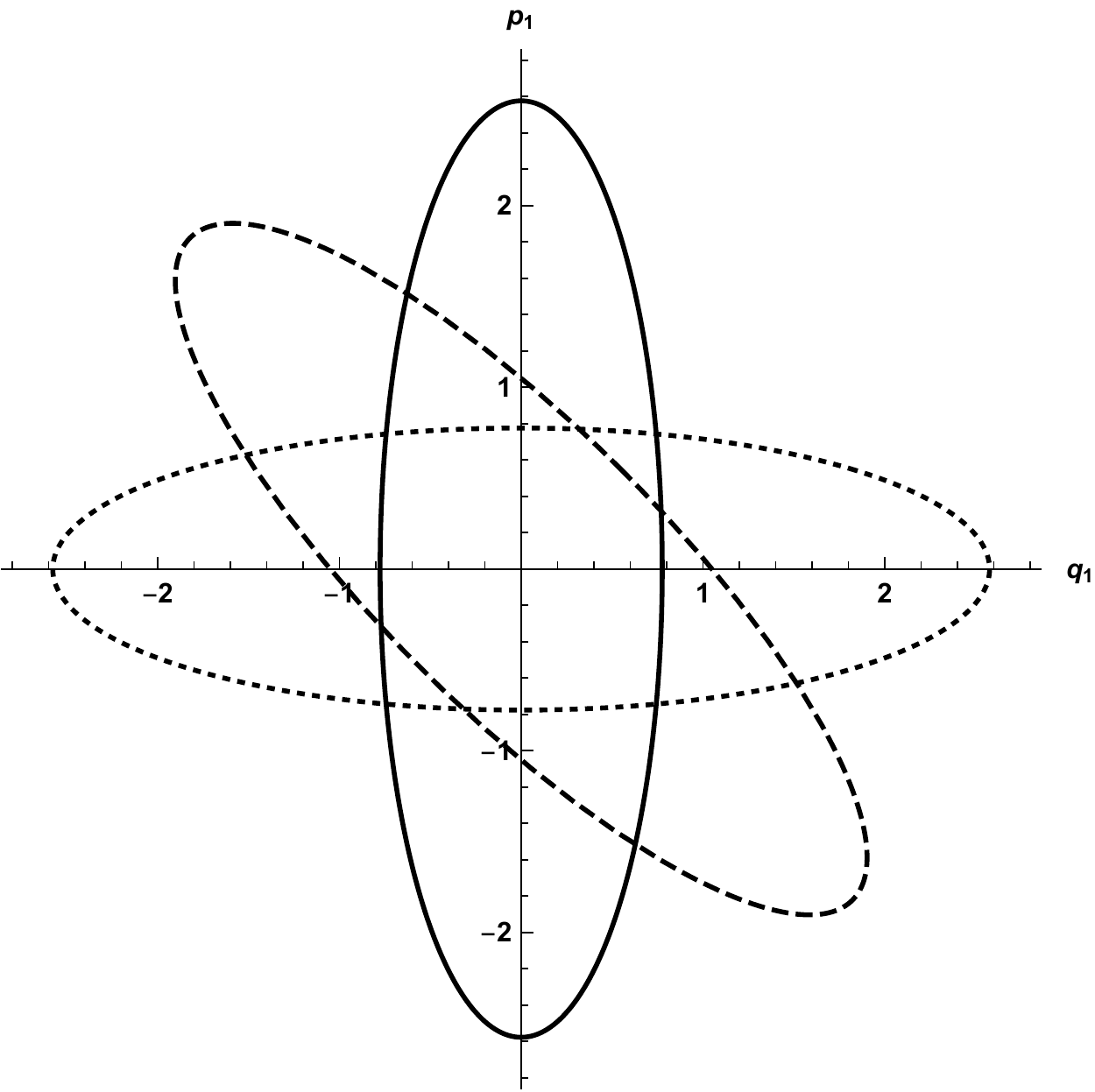}
         \caption{{\footnotesize{Rotating a squeezed trajectory with $r=0.6$. }}}
         \label{Rotation}
     \end{subfigure}
        \caption{{\footnotesize{ In both figures, the solid, the dashed and the dotted lines correspond to: (a) $r=0$, $r=0.3$ and $r=0.6$, respectively and (b) to $\phi=0$, $\phi=\pi/4$ and $\phi=\pi/2$, respectively.}}}
        \label{Figure}
\end{figure}

The action of ${\bf M}_{\bf s}$ on the trajectories is explicitly of the form
\begin{equation}
\left( \begin{array}{c} q'_1(t) \\ p'_1(t) \\ q'_2(t) \\ p'_2(t) \end{array} \right) = {\bf M}_{\bf s}\left(r, \phi \right) \left( \begin{array}{c} q_1(t) \\ p_1(t) \\ q_2(t) \\ p_2(t) \end{array} \right).
\end{equation}

In figure \ref{Figure} we showed the plot of $(q'_1(t), p'_1(t))$. As expected, we note in Fig. (\ref{Squeezing}), that the amount of squeezing $r$ squeezes the circular trajectory. Recall that symplectic transformations also preserve the area, hence the trajectories are squeezed but the area is preserved. On the other hand, the rotation angle $\phi$, as showed in (\ref{Rotation}), rotates the trajectories and also preserves the area.

Finally, observe that ${\bf M}_{\bf s}$ is given in the $\vec{Y}$ ``coordinatization'' which is not suitable for its quantum representation. To make it suitable, let us provide the expression for the matrix $\widetilde{\bf M}_{\bf s}$, which is given by
\begin{equation}
\widetilde{\bf M}_{\bf s} = \left(
\begin{array}{cccc}
 \cosh (r) & -\frac{l_1 \sinh (r) \cos (2 \phi )}{l_2} & 0 & -\frac{l_1 l_2 \sinh (r) \sin (2 \phi )}{\hbar } \\
 -\frac{l_2 \sinh (r) \cos (2 \phi )}{l_1} & \cosh (r) & -\frac{l_1 l_2 \sinh (r) \sin (2 \phi )}{\hbar } & 0 \\
 0 & -\frac{\hbar  \sinh (r) \sin (2 \phi )}{l_1 l_2} & \cosh (r) & \frac{l_2 \sinh (r) \cos (2 \phi )}{l_1} \\
 -\frac{\hbar  \sinh (r) \sin (2 \phi )}{l_1 l_2} & 0 & \frac{l_1 \sinh (r) \cos (2 \phi )}{l_2} & \cosh (r) \\
\end{array}
\right). \label{SqueezeMatrix}
\end{equation}

This expression will be used to explore the analog of the bipartite squeeze operator in polymer quantum mechanics in section \ref{LQCosmo}.

\subsection{Covariance matrix for squeezed states}

Now we will show the relation between the covariance matrix, denoted by ${\bf V}^{(2)}$, and the symplectic matrix $\widetilde{\bf M}$. Let us consider the state $| \Psi_{\widetilde{\bf M}} \rangle \in L^2(\mathbb{R}^n, d \vec{x})$ related with the symplectic matrix $\widetilde{\bf M}$ as
\begin{equation}
| \Psi_{\widetilde{\bf M}} \rangle = \widehat{C}_{\widetilde{\bf M}} | 0 \rangle , \label{InitialS}
\end{equation}
\noindent where $| 0 \rangle = \int d\vec{x} \; \Psi_0(\vec{x}) \, | \vec{x} \rangle $ is the state $|0\rangle = |0\rangle_1 \otimes | 0\rangle_2 \dots | 0 \rangle_n$, and the ket $| 0 \rangle_j$ is the vacuum state of the $j$-th quantum harmonic oscillator. Note that this construction can be extended to other states in $L^2(\mathbb{R}^n, d \vec{x})$ and not only for $| 0 \rangle$. However, for simplicity in our exposition, let us consider the simplest example of the covariance matrix for $ \widehat{C}_{\widetilde{\bf M}} | 0 \rangle$.

To obtain the covariance matrix we first calculate the following amplitude
\begin{equation}
\langle \Psi_{\widetilde{\bf M}} | \widehat{W}(\vec{a}, \vec{b}) | \Psi_{\widetilde{\bf M}} \rangle = \langle 0 | \widehat{C}^\dagger_{\widetilde{\bf M}} \; \widehat{W}(\vec{a}, \vec{b}) \; \widehat{C}_{\widetilde{\bf M}} | 0 \rangle, \label{Amplitude}
\end{equation}
\noindent where $\widehat{W}(\vec{a}, \vec{b})$ is the Weyl-algebra generator introduced in (\ref{WeylAlgGen}). Combining (\ref{UnitaryRep}), (\ref{PCRep}) and the Gaussian form of the vacuum state of the system given by $n$-decoupled harmonic oscillators, we obtain the following expression for the amplitude in  (\ref{Amplitude}) 
\begin{equation}
\langle \Psi_{\widetilde{\bf M}} | \widehat{W}(\vec{a}, \vec{b})  | \Psi_{\widetilde{\bf M}} \rangle  = \exp\left\{ - \frac{1}{4}   \left( \begin{array}{cc} \vec{a} & \vec{b} \end{array} \right)^T {\bf \Lambda} \left( \begin{array}{c} \vec{a} \\ \vec{b} \end{array}\right) \right\}, \label{Amplitude2}
\end{equation}
\noindent where the matrix ${\bf \Lambda}$ is given by
\begin{equation}
{\bf \Lambda} := {\widetilde{\bf M}} \left( \begin{array}{cc} \frac{1}{\hbar^2} {\bf L}^2 & {\bf 0} \\ {\bf 0} & {\bf L}^{-2} \end{array}\right) {\widetilde{\bf M}}^T, 
\end{equation}
\noindent and $ {\bf L} = \mbox{diag}(  l_1 , l_2 , \dots ,  l_n )$, where $l_j$ was defined earlier (\ref{AniCreOps}). 

The covariance matrix ${\bf V}^{(2)}$ has components given by
\begin{equation}
{\bf V}^{(2)} = \left( \begin{array}{cc} \langle \Psi_{\widetilde{\bf M}} | \widehat{x}_j \; \widehat{x}_k  | \Psi_{\widetilde{\bf M}} \rangle & \frac{1}{2}\langle \Psi_{\widetilde{\bf M}} |\left\{ \widehat{x}_j , \; \widehat{p}_k \right\} | \Psi_{\widetilde{\bf M}} \rangle \\  \frac{1}{2}\langle \Psi_{\widetilde{\bf M}} | \left\{ \widehat{p}_j, \; \widehat{x}_k \right\} | \Psi_{\widetilde{\bf M}} \rangle  & \langle \Psi_{\widetilde{\bf M}} | \widehat{p}_j \; \widehat{p}_k   | \Psi_{\widetilde{\bf M}} \rangle \end{array}\right),  \label{CVMatrixDef}
\end{equation}
\noindent and these components can be obtained from (\ref{Amplitude2}) using the following relations
\begin{eqnarray}
\langle \Psi_{\widetilde{\bf M}} | \widehat{x}_j \; \widehat{x}_k  | \Psi_{\widetilde{\bf M}} \rangle = -\hbar^2 \partial^2_{ a_j a_k} \langle \Psi_{\widetilde{\bf M}} | \widehat{W}(\vec{a}, \vec{b})  | \Psi_{\widetilde{\bf M}} \rangle \vert_{\vec{a}, \vec{b}=0}, \label{EqCV1}\\
\frac{1}{2}\langle \Psi_{\widetilde{\bf M}} | \left\{ \widehat{x}_j, \; \widehat{p}_k \right\} | \Psi_{\widetilde{\bf M}} \rangle = -\hbar^2  \partial^2_{a_j b_k} \langle \Psi_{\widetilde{\bf M}} | \widehat{W}(\vec{a}, \vec{b})  | \Psi_{\widetilde{\bf M}} \rangle \vert_{\vec{a}, \vec{b}=0},\label{EqCV2}  \\
\frac{1}{2}\langle \Psi_{\widetilde{\bf M}} | \left\{ \widehat{p}_j , \; \widehat{x}_k \right\} | \Psi_{\widetilde{\bf M}} \rangle = -\hbar^2  \partial^2_{b_j a_k} \langle \Psi_{\widetilde{\bf M}} |  \widehat{W}(\vec{a}, \vec{b})  | \Psi_{\widetilde{\bf M}} \rangle \vert_{\vec{a}, \vec{b}=0}, \label{EqCV3} \\
\langle \Psi_{\widetilde{\bf M}} |  \widehat{p}_j \; \widehat{p}_k  | \Psi_{\widetilde{\bf M}} \rangle = -\hbar^2 \partial^2_{b_j b_k} \langle \Psi_{\widetilde{\bf M}} |  \widehat{W}(\vec{a}, \vec{b})  | \Psi_{\widetilde{\bf M}} \rangle \vert_{\vec{a}, \vec{b}=0}. \label{EqCV4}
\end{eqnarray}
\noindent Remarkably, the resulting expression for ${\bf V}^{(2)} $ in terms of the symplectic matrix $\widetilde{\bf M}$ is 
\begin{equation}
{\bf V}^{(2)}  = \frac{1}{2}{\widetilde{\bf M}} \left( \begin{array}{cc} {\bf L}^2 & {\bf 0} \\ {\bf 0} & \hbar^2 {\bf L}^{-2} \end{array}\right) {\widetilde{\bf M}}^T, \label{CVMatrix}
\end{equation}
\noindent and this shows the direct relation between the covariance matrix ${\bf V}^{(2)}$ for the state $| \Psi_{\widetilde{\bf M}} \rangle$ and the symplectic matrix ${\widetilde{\bf M}}$ associated with the unitary operator $\widehat{C}_{\widetilde{\bf M}}$. Moreover, if we now consider the definition (\ref{1SpCond}), it can be shown that $\frac{2}{\hbar} {\bf V}^{(2)}$ is actually a symplectic matrix. Let us apply this formula to some of the systems considered before.

Consider the matrix ${\bf M}_1$ given in (\ref{Casebzero}). Using (\ref{RelationbetweenMs}) we obtain the expression for ${\widetilde{\bf M}}_1$ which then is replaced in (\ref{CVMatrix}) giving rise to the following covariance matrix
\begin{equation}
{\bf V}^{(2)}_1 = \frac{1}{2} \left(
\begin{array}{cccc}
{V}^{(2)}_{11} & 0 & {V}^{(2)}_{13} & 0 \\
0 & {V}^{(2)}_{22} & 0 & {V}^{(2)}_{24} \\
{V}^{(2)}_{13} & 0 & {V}^{(2)}_{33} & 0 \\
0 & {V}^{(2)}_{24} & 0 & {V}^{(2)}_{44}
\end{array}
\right).
\end{equation}
\noindent Its components are given in the appendix (\ref{V2MatrixCoeff}) and in the particular case where $a_{11} = a_{22} = c_{11} = c_{22} = 0$, the covariance matrix ${\bf V}^{(2)}_1$ reduces to
 \begin{equation}  \label{V2Squeeze}
{\bf V}^{(2)}_1 = \frac{1}{2} \left(
\begin{array}{cccc}
 l^2_1 e^{2 a_{12} } & 0 & 0 & 0 \\
0 & l^2_2 e^{2 c_{12} } & 0 & 0 \\
0 & 0 & \frac{ \hbar ^2 }{l^2_1} e^{ - 2 a_{12}}  & 0 \\
0 & 0 & 0 & \frac{ \hbar ^2 }{l^2_2} e^{ - 2 c_{12}}
\end{array}
\right).
\end{equation}

We use this result to derive the uncertainties in the coordinates $\Delta x_j$ for $j=1,2$ which according to (\ref{V2Squeeze}) are given by
\begin{equation} \label{xdesvsqeez1}
\Delta x_j := \sqrt{ \langle \Psi_{\widetilde{\bf M}} | \widehat{x}^2_j | \Psi_{\widetilde{\bf M}} \rangle - \langle \Psi_{\widetilde{\bf M}} | \widehat{x}_j | \Psi_{\widetilde{\bf M}} \rangle^2 } =  \frac{l_j e^{ \alpha_j }}{\sqrt{2}}  ,
\end{equation}
\noindent  where due to the symmetry of the vacuum wavefunction we have $\langle \Psi_{\widetilde{\bf M}} | \widehat{x}_j | \Psi_{\widetilde{\bf M}} \rangle =0$. This can be verified calculating the first derivatives in (\ref{Amplitude2}). Here, for simplicity we make $ \alpha_1 = a_{12}$ and $\alpha_2 = c_{12}$. Note the remarkably property of the squeezed states like in (\ref{InitialS}) which is that $\Delta x_j$ can be smaller than the proper length of the vacuum state $l_j$ when $\alpha_j < 0$.

 Similarly, the uncertainties in the momenta $\Delta p_j$, are
\begin{equation} \label{pdesvsqeez1}
\Delta p_j := \sqrt{ \langle \Psi_{\widetilde{\bf M}} | \widehat{p}^2_j | \Psi_{\widetilde{\bf M}} \rangle - \langle \Psi_{\widetilde{\bf M}} | \widehat{p}_j | \Psi_{\widetilde{\bf M}} \rangle^2 } =  \frac{\hbar}{ \sqrt{2} l_j e^{\alpha_j }}  ,
\end{equation} 
\noindent which can also be smaller than $ \frac{\hbar}{ l_j}$ when $\alpha_j > 0$ and also, $\langle \Psi_{\widetilde{\bf M}} | \widehat{p}_j | \Psi_{\widetilde{\bf M}} \rangle = 0$ as the previous case. Nevertheless, both uncertainties satisfy Heisenberg's uncertainty principle:
\begin{equation}
\Delta x_j \, \Delta p_j = \left( \frac{l_j e^{ \alpha_j }}{\sqrt{2}}\right) \left(  \frac{\hbar}{ \sqrt{2} l_j e^{\alpha_j }} \right) = \hbar/2. \label{HeisenbergUP}
\end{equation}

Another interesting covariance matrix is the one related with the bipartite squeeze operator (\ref{SqueezeMatrix}) derived in the previous subsection. Inserting (\ref{SqueezeMatrix}) in (\ref{CVMatrix}) yields
\begin{equation}
{\bf V}^{(2)}(r,\phi)  =\left( \begin{array}{cccc}
 \frac{l_1^2 \cosh (2 r)}{2}  & - \frac{l_1 l_2 \sinh (2r) \cos (2 \phi )}{2}  & 0 & -\frac{ l_1 \hbar  \sinh (2 r) \sin (2\phi ) }{ 2 l_2} \\
 - \frac{l_1 l_2 \sinh (2r)  \cos (2 \phi )}{2}  & \frac{l_2^2 \cosh (2 r)}{2}  & -\frac{ l_2 \hbar  \sinh (2r)  \sin (2 \phi ) }{2 l_1} & 0 \\
 0 & -\frac{l_2 \hbar  \sinh (2 r)  \sin (2 \phi )}{2 l_1} & \frac{\hbar ^2 \cosh (2 r)}{2 l_1^2} & \frac{\hbar ^2 \sinh (2 r)  \cos (2 \phi )}{2 l_1 l_2} \\
 -\frac{ l_1 \hbar  \sinh (2 r)  \sin (2 \phi )}{2 l_2} & 0 & \frac{\hbar ^2 \sinh (2 r) \cos (2 \phi )}{2 l_1 l_2} & \frac{\hbar ^2 \cosh (2 r)}{2 l_2^2} 
\end{array}
\right),
\end{equation}
\noindent and this allows us to determine the correlation between the second moments of the subsystem 1 and the subsystem 2
\begin{eqnarray}
\langle \Psi_{\widetilde{\bf M}} | \left(\widehat{x}_1 \pm  \widehat{x}_2 \right)^2 | \Psi_{\widetilde{\bf M}} \rangle &=& \frac{e^{2r}}{4}  \left[ l^2_1 + l^2_2 \mp 2 l_1 \, l_2 \cos(2 \phi) \right] + \frac{e^{-2r}}{4}  \left[ l^2_1 + l^2_2 \pm 2 l_1 \, l_2 \cos(2 \phi) \right] , \label{xdesvsqeez2} \\
\langle \Psi_{\widetilde{\bf M}} | \left(\widehat{p}_1 \pm  \widehat{p}_2 \right)^2 | \Psi_{\widetilde{\bf M}} \rangle &=& \frac{\hbar^2 e^{2r}}{4 l^2_1 \, l^2_2}  \left[ l^2_1 + l^2_2 \pm 2 l_1 \, l_2 \cos(2 \phi) \right] + \frac{\hbar^2 e^{-2r}}{4 l^2_1 \, l^2_2}  \left[ l^2_1 + l^2_2 \mp 2 l_1 \, l_2 \cos(2 \phi) \right] . \label{pdesvsqeez2}
\end{eqnarray}

In the particular case where $l_1 = l_2 = l$ and $\phi = \frac{\pi}{2}$, the uncertainties $ (\Delta x_1)_{\widetilde{\Psi}} $ and $ (\Delta x_2)_{\widetilde{\Psi}} $ for the state $\widetilde{\Psi}$ are correlated as follows
\begin{equation}
 (\Delta x_1)^2_{\widetilde{\Psi}} +  (\Delta x_2)^2_{\widetilde{\Psi}}  = l^2 \cosh(2r). \label{StandardCorr}
\end{equation}

These are the main results, at the standard quantum mechanics level, which we want to show regarding the representation of the symplectic group in quantum mechanics. There are others applications like the analysis of the Bohmian trajectories for bipartite squeezed states, the analysis of the fidelity for bipartite or tripartite squeezed states, and others which are currently in preparation. For now, let us move to the analysis of the squeezed states in polymer quantum mechanics given in the next section.

\section{Squeezed states in polymer quantum mechanics} \label{LQCosmo}

Polymer quantum mechanics \cite{ashtekar2003quantum, corichi2007polymer, velhinho2007quantum, pawlowski2014separable}, is a quantization scheme which can be considered as a ``toy model'' looming from loop quantum cosmology. Hence, exploring the nature and properties of squeezed states in polymer quantum mechanics will help study those scenarios in loop quantum cosmology where such states might play a significant role. 

For this example, we will consider a system with two degrees of freedom and both will be polymer quantized. Therefore, the Hilbert space of the entire system is given by 
\begin{equation}
{\cal H}_{poly} = {\cal H}^{(1)}_{poly} \times {\cal H}^{(2)}_{poly} ,
\end{equation}
\noindent where the Hilbert spaces ${\cal H}^{(j)}_{poly} $ with $j=1,2$ are of the form
\begin{equation}
{\cal H}^{(j)}_{poly} = L^2(\overline{\mathbb{R}}, dp^{(j)}_{Bohr}),
\end{equation}
\noindent  where $\overline{\mathbb{R}}$ is the Bohr compactification of real line and $dp^{(j)}_{Bohr}$ is the Bohr measure (see \cite{velhinho2007quantum} for more details). This Hilbert space resembles the momentum representation used in the standard quantum mechanics. 

An arbitrary state in this Hilbert space ${\cal H}_{poly}$ is given by
\begin{equation} \label{PlyState}
\Psi(p_1,p_2) = \sum_{\{ \vec{x}_j \}} \Psi_{\vec{x}_j} e^{ \frac{i}{\hbar}  \vec{x}^T_j \vec{p} },
\end{equation}
\noindent where $\{ \vec{x}_j \} $ is a shorthand notation for the graph  $\{ (x^{(1)}_j, x^{(2)}_j) \}^{j=n}_{j=1} $ associated with the state $\Psi(\vec{p})$. In this notation, the array $\vec{p} = \left( p_1, p_2 \right)$ denotes the momentum variables for the system 1 and 2, respectively. The coefficients $\Psi_{\vec{x}_j}$ provide the value for the norm of the state which is given by
\begin{equation}
|| \Psi(\vec{p}) || = \sum_{ \{ \vec{x}_j \} } |\Psi_{\vec{x}_j}|^2,
\end{equation}
\noindent hence these coefficients are different from zero and the sum converges (they are non-null over countable number of points in the graph $\{ \vec{x}_j \}$). This norm arises from the inner product
\begin{equation}
\langle \Psi | \Phi \rangle  = \lim_{L_1, L_2 \rightarrow \infty} \frac{1}{4 \, L_1 L_2} \int^{L_1}_{- L_1} \int^{L_2}_{- L_2} \Psi^*(\vec{p}) \, \Phi(\vec{p}) dp_1 dp_2 ,
\end{equation}
\noindent which for the specific case of the plane waves takes the form of the Kronecker delta
\begin{equation}
\langle e^{\frac{i}{\hbar} \vec{x}^T \vec{p}} | e^{\frac{i}{\hbar} \vec{x}'^T \vec{p}} \rangle = \delta_{\vec{x}, \vec{x}'}.
\end{equation}

This inner product is the main signature of the polymer quantization as it violates the Stone-von Neumann theorem. Consequently, polymer quantum mechanics is not unitarily equivalent to the standard Schr\"odinger representation. Moreover, in polymer quantum mechanics there is no momentum operator hence infinitesimal spatial translations cannot be implemented. Nevertheless, we can obtain a representation for the position operator, which in the present case is given by
\begin{eqnarray} 
\widehat{q}_1 \Psi(p_1, p_2) &=& i \hbar \frac{\partial }{\partial p_1} \Psi(p_1, p_2) = -  \sum_{ \{ \vec{x}_j \} } \Psi_{\vec{x}_j}  x^{(1)}_j e^{ \frac{i}{\hbar} \vec{x}^T_j \vec{p}}, \label{PositionRep1} \\
\widehat{q}_2 \Psi(p_1, p_2) &=& i \hbar \frac{\partial }{\partial p_2} \Psi(p_1, p_2) = - \sum_{ \{ \vec{x}_j \} } \Psi_{\vec{x}_j}  x^{(2)}_j e^{ \frac{i}{\hbar} \vec{x}^T_j \vec{p}}. \label{PositionRep2}
\end{eqnarray} 

Despite these peculiarities with the non-regularity of the polymer representation, the representation of the symplectic group $Sp(2n,\mathbb{R})$ on the Hilbert space of polymer quantum mechanics was provided recently by one of the authors in Ref. \cite{Garcia-Chung:2020cag}. There, the representation is given by the map $ \widehat{C}^{(poly)} : Sp(2n, \mathbb{R}) \rightarrow L({\cal H}_{poly}),\; \widetilde{\bf M} \mapsto \widehat{C}^{(poly)}_{\widetilde{\bf M}}$, where the linear operator $\widehat{C}^{(poly)}_{\widetilde{\bf M}}$ acts on ${\cal H}_{poly}$ as
\begin{equation} \label{SymPolyRep}
\widehat{C}^{(poly)}_{\widetilde{\bf M}} \Psi(\vec{p})= \lim_{L_1, L_2  \rightarrow \infty} \frac{1}{4 \, L_1 \, L_2}\int^{L_1}_{-L_1} \int^{L_2}_{-L_2}  C^{(poly)}_{\widetilde{\bf M}}(\vec{p}, \vec{p}') \Psi(\vec{p}') d\vec{p}'.
\end{equation}
The polymer kernel $C^{(poly)}_{\widetilde{\bf M}}(\vec{p}, \vec{p}')$ is given by
\begin{equation}
C^{(poly)}_{\widetilde{\bf M}}(\vec{p}, \vec{p}') = \det (\widetilde{\bf D} \widetilde{\bf A}^T )^{-\frac{1}{4}} e^{ -\frac{i}{2 \hbar} \vec{p}^T  \widetilde{\bf B} \widetilde{\bf D}^{-1} \vec{p}} \sum_{\vec{x}} e^{ \frac{i}{\hbar} \vec{p}^T \vec{x} - \frac{i}{\hbar} \vec{p}'^T \widetilde{\bf D}^T \vec{x} + \frac{i}{2\hbar} \vec{x}^T {  \widetilde{ \bf D} \widetilde{\bf C}^T  } \vec{x}}, \label{KSpPoly}
\end{equation}
\noindent and note that when the factor $\det (\widetilde{\bf D} \widetilde{\bf A}^T )^{-\frac{1}{4}}\neq 1$ it implies that this representation is not unitary \cite{Garcia-Chung:2020cag}.

Recall that one of the main features of loop quantum cosmology is its intrinsic length scale given by the Planck length. As a toy model, polymer quantum mechanics does not have an intrinsic length scale. However, it admits a length scale that mimics some of the features of loop quantum cosmology. This length scale is introduced at hand and is called {\it polymer scale}, usually denoted by $\mu$. This polymer scale constitutes the analog of minimum length for polymer quantum mechanics models, and therefore, it can be considered as a lower bound for the uncertainties. In the present analysis, each system admits a polymer scale $\mu_1$ and $\mu _2$ when the dynamics is considered. 

Let us now consider the following questions: (1) is it possible to have polymer states such that their uncertainties are lower than the polymer scale? Furthermore, (2) do the correlations found in (\ref{xdesvsqeez2}) have an analog in polymer quantum mechanics?

To answer these questions let us consider the matrix $\widetilde{\bf M}_1$ with $a_{11} = a_{22} = c_{11} = c_{22} = 0$, used to calculate the covariance matrix ${\bf V}^{(2)}_1$ in (\ref{V2Squeeze}), but now with $a_{12} = - r_1$ and $c_{12} = - r_2$. The explicit form is $\widetilde{\bf M}_1$
\begin{equation}
\widetilde{\bf M}_1 =  \left( \begin{array}{cccc} e^{-r_1 } & 0 & 0 & 0 \\ 0 & e^{-r_2} & 0 & 0 \\ 0 & 0 &  e^{r_1} & 0 \\ 0 & 0 & 0 & e^{r_2} \end{array}\right). \label{Casebzerotilde}
\end{equation}

The action of the group element $\widehat{C}^{(poly)}_{\widetilde{\bf M}_1}$ on an arbitrary polymer state (\ref{PlyState}) gives the following state
\begin{equation}
\widetilde{\Psi}_{\widetilde{\bf M}_1}(p_1, p_2) = \widehat{C}^{(poly)}_{\widetilde{\bf M}_1} \Psi(p_1, p_2) = \sum_{ \{ \vec{x}_j \} } \Psi_{\vec{x}_j} e^{ \frac{i}{\hbar} \left( e^{-r_1} p_1 x^{(1)}_j + e^{-r_2} p_2 x^{(2)}_j\right) }.  \label{SqeezedPolymerState1}
\end{equation}

Using the representation of the position operators $\widehat{q}_1$ and $\widehat{q}_2$ given in (\ref{PositionRep1}) and (\ref{PositionRep2}) we obtain the dispersion relations
\begin{equation} \label{PolymerDispersionRel}
(\Delta x_1)_{\widetilde{\Psi}} = e^{-r_1} ( \Delta x_1)_{\Psi} , \qquad (\Delta x_2)_{\widetilde{\Psi}} = e^{-r_2} ( \Delta x_2)_{\Psi} ,
\end{equation}
\noindent which show that the squeezed polymer state is indeed squeezed by a factor $e^{-r_1}$ or $e^{-r_2}$. Consequently, if the initial dispersion of the polymer state is given by $( \Delta x_1)_{\Psi}$ or $( \Delta x_2)_{\Psi} $, then the squeeze operator $\widehat{C}^{(poly)}_{\widetilde{\bf M}_1}$ gives rise to a polymer state $\widetilde{\Psi}$ (\ref{SqeezedPolymerState1}) whose dispersion is smaller than that of the initial polymer state $\Psi$. Moreover, due to there is no upper bound for the parameter $r$, these dispersion relations can be smaller than the corresponding polymer scales $\mu_1$ and $\mu_2$. 

Let us now consider the analog of the correlations (\ref{StandardCorr}) but for polymer states. To do so, consider the polymer representation $\widehat{C}^{(poly)}_{\widetilde{\bf M}_{\bf s}}$ of the symplectic matrix (\ref{SqueezeMatrix}) corresponding to a bi-partite system. The action of $\widehat{C}^{(poly)}_{\widetilde{\bf M}_{\bf s}}$ on an arbitrary polymer state (\ref{PlyState}) is given by
\begin{equation}
\widetilde{\Psi}_{s} = \widehat{C}^{(poly)}_{\widetilde{\bf M}_{\bf s}} \Psi(p_1,p_2) = \sum_{ \{ \vec{x}_j \} } \Psi_{\vec{x_j}}  e^{ \frac{i}{\hbar} \left[ \left(  \cosh(r) \, x^{(1)}_j  + \sinh(r) \, x^{(2)}_j \right) p_1 + \left(  \cosh(r) \, x^{(2)}_j  + \sinh(r) \, x^{(1)}_j \right) p_2 \right] }, \label{SqeezedPolymerState2}
\end{equation}
\noindent where again the representation in (\ref{SymPolyRep}) was used.

We combine this result with the representation of the position operators in (\ref{PositionRep1}) and (\ref{PositionRep2}) and obtain the following relations
\begin{equation} \label{PolyDispRel1}
(\Delta x_1)^2_{\widetilde{\Psi}_s} - (\Delta x_2)^2_{\widetilde{\Psi}_s} = (\Delta x_1)^2_{\Psi} - (\Delta x_2)^2_{\Psi},  \qquad \forall \quad \Psi \in {\cal H}_{poly}  ,
\end{equation}
\noindent where the conditions $l_1 = l_2$ and $\phi = \{ 0, \frac{\pi}{2} , \pi \}$ were imposed. Remarkably, this result not only is independent of the parameter $r$ (which labels the amount of squeezing) but also applies to any polymer state $\Psi \in {\cal H}_{poly}$. As can be seen, the difference of dispersions squared is conserved, regardless of the amount of squeezing. Also, note that $l_1$ and $l_2$ are considered group parameters and have no relation to the dynamics, i.e., we are considering general states in ${\cal H}_{poly}$. The same applies for $\phi$.

Let us now consider pure and symmetric polymeric states. The pure states are those that can be written as the following product
\begin{equation}
\Psi^{(p)}(p_1, p_2) = \left( \sum_{ \{ x^{(1)}_j \} } \Psi^{(1)}_{x^{(1)}_j } e^{ \frac{i}{\hbar} p_1 x^{(1)}_j } \right) \left( \sum_{ \{ x^{(2)}_j \} } \Psi^{(2)}_{x^{(2)}_j } e^{ \frac{i}{\hbar} p_2 x^{(2)}_j } \right).
\end{equation}
\noindent Secondly, both lattices $\{ x^{(1)}_j \}$ and $\{ x^{(2)}_j \}$ are symmetric, i.e., for every positive point $0 < x^{(s)}_j \in \{ x^{(s)}_j \} $ there exist a negative point $ 0 > x^{(s)}_{j'} \in \{ x^{(s)}_j \}$, such that $x^{(s)}_j + x^{(s)}_{j'} =0 $, and the states are also symmetric which implies that $\Psi^{(s)}_{ x^{(s)}_j } = \Psi^{(s)}_{ x^{(s)}_{j'} } $. These states are the analog of the states described with even functions in the standard quantum mechanics. 

The dispersion relation for squeezed pure symmetric polymer states is given by
\begin{equation}
(\Delta x_1)^2_{\widetilde{\Psi}_s} + (\Delta x_2)^2_{\widetilde{\Psi}_s} = \cosh(2r) \left( (\Delta x_1)^2_{\Psi^{(p)}} + (\Delta x_2)^2_{\Psi^{(p)}} \right),
\end{equation}
\noindent which takes the form
\begin{equation}
(\Delta x_1)^2_{\widetilde{\Psi}_s} + (\Delta x_2)^2_{\widetilde{\Psi}_s} = l^2 \cosh(2r) , \label{PolyDispRel2}
\end{equation}
\noindent when the dispersion of the pure states are $\left( (\Delta x_1)_{\Psi^{(p)}} = (\Delta x_2)_{\Psi^{(p)}} \right) = l/ \sqrt{2}$. 
Notably, Eq. (\ref{PolyDispRel2}) is the same as that obtained in (\ref{StandardCorr}) for the Schr\"odinger representation. This shows that the correlations present in the standard quantum mechanics using the $\widehat{C}_{\widetilde{\bf M}}$ operator for both symplectic matrices $\widetilde{\bf M}_1$ and $\widetilde{\bf M}_{\bf s}$ are the same to those obtained in polymer quantum mechanics using the operator $\widehat{C}^{(poly)}_{\widetilde{\bf M}}$.

\section{Conclusions} \label{Conclusion}

In this paper we provided the direct relation between the Lie algebra $sp(4, \mathbb{R})$ and the symplectic group $Sp(4, \mathbb{R})$. The expression shows the link between the block matrices ${\bf A}$, ${\bf B}$, ${\bf C}$ and ${\bf D}$ with those of the Lie algebra ${\bf a}$, ${\bf b}$ and ${\bf c}$ given in the Eqs. (\ref{TerminoA})-(\ref{TerminoD}). This result has not been reported before and applies to the full Lie algebra $sp(4, \mathbb{R})$ of the symplectic group $Sp(4, \mathbb{R})$. 

Such relation allows us to obtain some important symplectic matrices that were used in subsequent sections. In the first example for ${\bf a}, {\bf c} \neq {\bf 0}$ and ${\bf b} =0$, we show that the corresponding symplectic matrix ${\bf M}_1$, given in (\ref{Casebzero}), can be written as $\left( {\bf M}'_1 \otimes {\bf 1} \right) \cdot \left( {\bf 1} \otimes {\bf M}'_2 \right)$, where ${\bf 1}, {\bf M}'_1, {\bf M}'_2 \in Sp(2, \mathbb{R})$. Here, ${\bf M}'_1$ and ${\bf M}'_2$ are symplectic matrices acting over each of the sub-systems with coordinates $(q_1, p_1)$ and $(q_2, p_2)$ respectively. In the case ${\bf a} = {\bf c} = \mbox{diag}(a_{11}, a_{22})$ and ${\bf b} =\mbox{diag}(b_{11}, b_{22})$ the symplectic matrix ${\bf M}_2$ in (\ref{SecondExample}), describes two coupled harmonic oscillators with interaction terms labeled by the coefficients of the matrix ${\bf b}$. Finally, in the equation (\ref{Caseaandczero}) we showed the symplectic matrix ${\bf M}_3$ for the case in which ${\bf a} = {\bf c} = {\bf 0}$ and a general form of the matrix ${\bf b}$.

In section (\ref{QUANTREP}) we analyzed the classical description of squeeze operators. We showed that the symplectic matrix ${\bf M}_{\bf s}$ is the classical analog of the squeeze operator ${\widehat{S}(\zeta) = e^{\widehat{s}_{\zeta}}} = \widehat{C}_{\widetilde{\bf M}_{\bf s}}$. Also, we remarked the isomorphism between the Lie algebra $sp(4,\mathbb{R})$ and ${\cal P}(2, \mathbb{R})$. Additionally, the general form of the covariance matrix ${\bf V}^{(2)}$ for the squeezed vacuum state $| \Psi_{\widetilde{\bf M}} \rangle$ was derived using the Weyl algebra representation and the symplectic matrix ${\bf M}_1$. The components of this covariant matrix were used to calculate the dispersion relations (\ref{xdesvsqeez1}) and (\ref{pdesvsqeez1}) for the particular case where $a_{11} =a_{22} = c_{11} = c_{22} = 0$. As is already known, these dispersions can be smaller than the vacuum characteristic length for the harmonic oscillators. They also satisfy the Heisenberg uncertainty principle as was shown in (\ref{HeisenbergUP}). We then calculated the covariance matrix for the symplectic matrix ${\bf M}_{\bf s}$ corresponding to the classical analog of the bipartite squeeze operator. With this matrix we determined the correlation (\ref{StandardCorr}). We also provided the general expressions for these correlations in equations (\ref{xdesvsqeez2}) and (\ref{pdesvsqeez2}).

Applying the previous results, it is also possible to represent operators in non-regular Hilbert spaces that are non-unitarily equivalent to the Fock-Schr\"odinger representation, so in section (\ref{LQCosmo}) we analyze polymer quantized systems. We calculated the dispersion relation for an arbitrary polymer state using the representation of the symplectic group in polymer quantum mechanics. We obtained that the polymer representation of the squeeze operator given by $\widehat{C}^{(poly)}_{\widetilde{\bf M}_1}$, yields a dispersion relation Eq. (\ref{PolymerDispersionRel}), which can be smaller than those of the initial state. This implies that $\widehat{C}^{(poly)}_{\widetilde{\bf M}_1}$ is indeed a polymer squeeze operator and (\ref{SqeezedPolymerState1}) describes a polymer squeezed state.

On the other hand, the polymer representation of the bipartite squeeze operator given by $\widehat{C}^{(poly)}_{\widetilde{\bf M}_{\bf s}}$ was used to derive the polymer correlations (\ref{PolyDispRel1}) and (\ref{PolyDispRel2}). The first correlation (\ref{PolyDispRel1}) shows that the difference of the dispersions' square is preserved and is independent of the initial polymer state. Clearly, this result only holds for $\widehat{C}^{(poly)}_{\widetilde{\bf M}_{\bf s}}$ so a symplectic matrix different  than the one used in (\ref{Casebzerotilde}) will produce a different result. In the case of (\ref{PolyDispRel2}), the result has the same form as the standard correlation in Eq. (\ref{StandardCorr}), hence, the polymer representation of $\widehat{C}^{(poly)}_{\widetilde{\bf M}_{\bf s}}$ can be used to construct correlated squeezed states for bipartite polymer systems. Naturally, this brings some questions like whether there is any mechanism in nature, say loop quantum cosmology or the interior of a black hole, from which a polymer squeezed state can be created.

Moreover, establishing that squeezing is a property also present in non-regular representations questions its role in the classicality of some cosmological models. As we showed, it is possible to construct entangled polymer states using $\widehat{C}^{(poly)}_{\widetilde {\bf M}_{\bf s}}$. Such polymer entangled states correlations satisfy a relation identical to the one obtained in the standard quantum mechanics. In this case, the states are polymer bipartite squeezed states, similar to those used in the quantum description of the inflaton field.

Finally, it is worth to mention that the polymer squeezed states obtained as a result of the representation of $Sp(4,\mathbb{R})$ in ${\cal H}_{poly}$, given in (\ref{SqeezedPolymerState1}, \ref{SqeezedPolymerState2}), differ from those reported in the LQC literature \cite{taveras2008corrections, mielczarek2012gaussian, gazeau2013quantum, diener2014numerical, diener2014numerical2, corichi2011coherent}. In these references, a Gaussian form of the states is considered, whereas in our case, the polymer state (\ref{SqeezedPolymerState1}) is a general polymer state. In LQC, the squeezed states are constructed by hand due after imposing some conditions to achieve the squeezed nature of the dispersion relations. In our case, the squeezed state results from the action of the squeeze operator. 

The results of this work open the doors for studying the entangled states of matter and geometry and the role that their correlations might play in some physical scenarios.

\section{Acknowledgments}
I thank Academia de Matem\'aticas and Colegio de F\'isica, UP, for the support and enthusiasm.

\section{Appendix: calculation of ${\bf m}^{2n}$} \label{Appendix1}
In this appendix the expression for the matrix ${\bf m}^{2n}$ is obtained. To do so, recall that the matrix ${\bf m}^2$ is formed by four $2 \times 2$ block matrices where the upper left and the lower right matrices are multiples of the identity matrix ${\bf 1}_{2\times2}$. The upper right block is the matrix ${\bf J} {\bf d}$ whereas the lower left is $- {\bf J} {\bf d}^T$. Notably, we found that this block structure is preserved after exponentiating the matrix ${\bf m}^2$ an integer number of times. That is to say, the n-power of matrix ${\bf m}^2$ yields a new matrix $({\bf m}^{2})^n$ given by
\begin{equation}
({\bf m}^{2})^n = \left( \begin{array}{cc} \alpha_n{\bf 1}_{2\times2} & \beta_n {\bf J} {\bf d} \\ - \beta_n {\bf J} {\bf d}^T & \gamma_n {\bf 1}_{2\times2} \end{array}\right). \label{MatrixSDef}
\end{equation} 
\noindent It is this pattern the one to be considered when this procedure is applied to higher order symplectic groups. 

The coefficients $\alpha_n$, $\beta_n$ and $\gamma_n$, are to be determined and depend on the values of the matrices ${\bf a}$, ${\bf b}$, ${\bf c}$ and ${\bf d}$. For $n=1$, these coefficients are given by the factors in the block matrices of ${\bf m}^2 $ given in  (\ref{SDefinition}) and can be directly defined as
\begin{equation}
\alpha_1 := - (\det {\bf a} + \det {\bf b}), \qquad \beta_1 := +1, \qquad \gamma_1 := - (\det {\bf c} + \det {\bf b}). \label{initialvalues}
\end{equation}

To calculate $\alpha_n$, $\beta_n$ and $\gamma_n$ for arbitrary $n$, first note that they can be generated with the $(n-1)$-power of the matrix ${\bf T}$ as
\begin{eqnarray}
\left( \begin{array}{c} \alpha_n \\ \beta_n \\ \gamma_n \end{array}\right) = {\bf T}^{n-1} \left( \begin{array}{c} \alpha_1 \\ \beta_1 \\ \gamma_1 \end{array}\right), \label{EqForCoeff}
\end{eqnarray}
\noindent where the matrix ${\bf T}$ is given by
\begin{equation}
{\bf T} = \left( \begin{array}{ccc} \alpha_1 & \beta_1 \det {\bf d} & 0 \\ \beta_1 & \gamma_1 & 0 \\ 0 & \beta_1 \det {\bf d} & \gamma_1 \end{array}\right).
\end{equation}

The calculation shows that ${\bf T}^{n-1}$ is a matrix of the form
\begin{equation}
{\bf T}^{n-1} = \left( \begin{array}{cc} {\bf U}^{n-1} & \vec{0}^T \\ \vec{u}^T \gamma_1^{n-2} \sum^{n-2}_{j=0} \gamma^{-j}_1 {\bf U}^j & \gamma^{n-1}_1 \end{array}\right),
\end{equation}
\noindent where $\vec{0} = (0,0)$ and $\vec{u} = (0,  \beta_1 \det {\bf d})$ and matrix ${\bf U}$ is given by
\begin{eqnarray}
{\bf U} = \left( \begin{array}{cc} \alpha_1 & \beta_1 \det {\bf d} \\ \beta_1 & \gamma_1\end{array}\right).
\end{eqnarray}

Then, using (\ref{EqForCoeff}) we have the following relation for the coefficients
\begin{eqnarray}
\left( \begin{array}{c} \alpha_n \\ \beta_n \end{array}\right) = {\bf U}^{n-1} \left( \begin{array}{c} \alpha_1 \\ \beta_1 \end{array}\right), \qquad \gamma_n = \gamma^n_1 + \vec{u}^T \gamma^{n-2}_1 \sum^{n-2}_{j=0} \gamma^{-j}_1 {\bf U}^j \left( \begin{array}{c} \alpha_1 \\ \beta_1 \end{array}\right). \label{FExpressionCoeff}
\end{eqnarray}

In order to calculate ${\bf U}^{n-1}$ we need to diagonalize the matrix ${\bf U}$. Let ${\bf P}$ be the matrix diagonalizing ${\bf U}$, then 
\begin{equation}
{\bf U} = {\bf P} \, {\bf D}_0 \, {\bf P}^{-1}, \label{EqForU}
\end{equation}
\noindent where the matrix ${\bf P}$ is
\begin{equation}
{\bf P} = \left( \begin{array}{cc} \frac{(\lambda_+ - \gamma_1)}{\beta_1 } k_1 & \frac{(\lambda_{-} - \gamma_1)}{\beta_1} k_2 \\ k_1 & k_2 \end{array}\right).
\end{equation}

The real arbitrary parameters $k_1$ and $k_2$ result from the diagonalization procedure. Its values will be automatically cancelled as part of the calculation of ${\bf U}^{n-1}$ further below. The eigenvalues of ${\bf U}$, denoted by $\lambda_{\pm}$, have the following expression 
\begin{eqnarray}
\lambda_{\pm} &=& \frac{\alpha_1 + \gamma_1}{2} \pm \frac{1}{2} \sqrt{ (\alpha_1 - \gamma_1)^2 + 4  \beta^2_1 \det {\bf d}}, \nonumber \\
 &=&  - \frac{\det {\bf a}  + \det {\bf c} + 2 \det {\bf b} }{2} \pm \frac{1}{2} \sqrt{ (\det {\bf a} - \det {\bf c})^2 + 4  \det {\bf d} }. \label{eigenvalues}
\end{eqnarray}
\noindent and the diagonal matrix ${\bf D}_0$ is
\begin{equation}
{\bf D}_0 = \left( \begin{array}{cc} \lambda_+ & 0 \\ 0 & \lambda_{-} \end{array}\right).
\end{equation}

We now take the $n-1$ power of ${\bf U}$ given in (\ref{EqForU}) to obtain the following result
\begin{eqnarray}
{\bf U}^{n-1} = \left( \begin{array}{cc} \frac{(\lambda_+ - \gamma_1)}{\beta_1 } k_1 & \frac{(\lambda_{-} - \gamma_1)}{\beta_1} k_2 \\ k_1 & k_2 \end{array}\right) \left( \begin{array}{cc} \lambda^{n-1}_+ & 0 \\ 0 & \lambda^{n-1}_{-} \end{array}\right) \left( \begin{array}{cc} \frac{(\lambda_+ - \gamma_1)}{\beta_1 } k_1 & \frac{(\lambda_{-} - \gamma_1)}{\beta_1} k_2 \\ k_1 & k_2 \end{array}\right)^{-1},
\end{eqnarray}
\noindent which, when combined with the result in (\ref{FExpressionCoeff}) together with the expression for $\vec{u}$, gives
\begin{eqnarray}
\alpha_n &=& \frac{ ( \lambda_+ - \det {\bf b}  - \det {\bf c}  ) \lambda^n_+ - ( \lambda_- - \det {\bf b}  - \det {\bf c}  ) \lambda^n_-   }{\sqrt{ (\det {\bf a} - \det {\bf c})^2 + 4 \beta^2_1 \det {\bf d} } }  ,\label{alpha} \\
\beta_n &=& \frac{ \lambda^n_+ -  \lambda^n_- }{\sqrt{ (\det {\bf a} - \det {\bf c})^2 + 4 \beta^2_1 \det {\bf d} } }  ,\label{beta} \\
\gamma_n &=& \frac{\left[ ( \lambda_+ - \det {\bf b}  - \det {\bf c} ) \lambda^n_- - ( \lambda_- - \det {\bf b}  - \det {\bf c} ) \lambda^n_+   \right]}{\sqrt{ (\det {\bf a} - \det {\bf c})^2 + 4 \beta^2_1 \det {\bf d} } } . \label{gamma}
\end{eqnarray}
\noindent These are the final expressions for the coefficients in $({\bf m}^{2})^n$.

\section{Appendix: Series analysis} \label{Series}

In this appendix we calculate the series expansion terms. To do so, note that once the expression for $({\bf m}^{2})^n$ is inserted the expansion (\ref{Expansion}) and  the even and odd terms are collected, the matrix ${\bf M}({\bf a}, {\bf b}, {\bf c})$ takes the form
\begin{equation}
{\bf M}({\bf a}, {\bf b}, {\bf c}) = \left( \begin{array}{cc}  \alpha^{(e)} {\bf 1}_{2\times2} &  \beta^{(e)} {\bf J} {\bf d} \\ - \beta^{(e)} {\bf J} {\bf d}^T & \gamma^{(e)} {\bf 1}_{2\times2} \end{array}\right) + {\bf m}  \left( \begin{array}{cc}  \alpha^{(o)} {\bf 1}_{2 \times 2} & \beta^{(o)} {\bf J} {\bf d} \\ - \beta^{(o)} {\bf J} {\bf d}^T &  \gamma^{(o)} {\bf 1}_{2 \times 2} \end{array}\right) , \label{FinalformM}
\end{equation}
\noindent where the following coefficients are given by
\begin{equation}
\alpha^{(e)} := 1 + \sum^{+\infty}_{n=1} \frac{1}{(2n)!} \alpha_n, \qquad \beta^{(e)} := \sum^{+\infty}_{n=1} \frac{1}{(2n)!} \beta_n, \qquad \gamma^{(e)} := 1 + \sum^{+\infty}_{n=1} \frac{1}{(2n)!} \gamma_n, \label{EvenCoeff}
\end{equation}
\begin{equation}
\alpha^{(o)} := 1 + \sum^{+\infty}_{n=1} \frac{1}{(2n+1)!} \alpha_n, \qquad \beta^{(o)} := \sum^{+\infty}_{n=1} \frac{1}{(2n+1)!} \beta_n, \qquad \gamma^{(o)} := 1 + \sum^{+\infty}_{n=1} \frac{1}{(2n+1)!}\gamma_n. \label{OddCoeff}
\end{equation}

We now insert (\ref{alpha}), (\ref{beta}) and (\ref{gamma}) in the relations (\ref{EvenCoeff}) - (\ref{OddCoeff}) to obtain
\begin{equation}
 \alpha^{(e)} =  \frac{1}{2} \left[ \cosh\left( \sqrt{\lambda_+} \right) + \cosh\left( \sqrt{\lambda_-} \right)\right]  +  \frac{\left( \det{\bf c} - \det{\bf a}\right) \left[ \cosh\left( \sqrt{\lambda_+} \right) - \cosh\left( \sqrt{\lambda_-} \right) \right] }{ 2 \sqrt{ (\det{\bf a} - \det{\bf c})^2 + 4 \det {\bf d} } }  , \label{alphapar} 
 \end{equation}
 
 \begin{eqnarray}
 \alpha^{(o)} = \frac{1}{2} \left[ \frac{ \sinh\left( \sqrt{\lambda_+} \right)}{ \sqrt{\lambda_+} } + \frac{ \sinh\left( \sqrt{\lambda_-} \right) }{ \sqrt{\lambda_-}}  \right] +  \frac{\left( \det{\bf c} - \det{\bf a}\right)  }{ 2  \sqrt{ (\det{\bf a} - \det{\bf c})^2 + 4 \det {\bf d} } } \left[ \frac{ \sinh\left( \sqrt{\lambda_+} \right)}{ \sqrt{\lambda_+} } - \frac{ \sinh\left( \sqrt{\lambda_-} \right) }{ \sqrt{\lambda_-}}  \right] , \nonumber \\ \label{alphaimpar}  
 \end{eqnarray}
 \begin{equation}
\beta^{(e)} = \frac{1}{\sqrt{ (\det{\bf a} - \det{\bf c})^2 + 4 \det {\bf d}  } } \left[  \cosh\left( \sqrt{\lambda_+}\right) -  \cosh\left(\sqrt{\lambda_- }\right) \right] , \label{betapar} 
\end{equation}
\begin{equation}
\beta^{(o)} = \frac{1}{  \sqrt{ (\det{\bf a} - \det{\bf c})^2 + 4 \det {\bf d} } } \left[ \frac{ \sinh\left( \sqrt{\lambda_+} \right)}{ \sqrt{\lambda_+} } - \frac{ \sinh\left( \sqrt{\lambda_-} \right) }{ \sqrt{\lambda_-}}  \right] , \label{betaimpar} 
\end{equation}

\begin{equation}
 \gamma^{(e)} =  \frac{1}{2} \left[ \cosh\left( \sqrt{\lambda_+} \right) + \cosh\left( \sqrt{\lambda_-} \right)\right]  -  \frac{\left( \det{\bf c} - \det{\bf a}\right) \left[ \cosh\left( \sqrt{\lambda_+} \right) - \cosh\left( \sqrt{\lambda_-} \right) \right] }{ 2 \sqrt{ (\det{\bf a} - \det{\bf c})^2 + 4 \det {\bf d} } }   , \label{gammapar} 
 \end{equation}
 
 \begin{eqnarray}
 \gamma^{(o)} =\frac{1}{2} \left[ \frac{ \sinh\left( \sqrt{\lambda_+} \right)}{ \sqrt{\lambda_+} } + \frac{ \sinh\left( \sqrt{\lambda_-} \right) }{ \sqrt{\lambda_-}}  \right] -  \frac{\left( \det{\bf c} - \det{\bf a}\right)  }{ 2  \sqrt{ (\det{\bf a} - \det{\bf c})^2 + 4 \det {\bf d} } } \left[ \frac{ \sinh\left( \sqrt{\lambda_+} \right)}{ \sqrt{\lambda_+} } - \frac{ \sinh\left( \sqrt{\lambda_-} \right) }{ \sqrt{\lambda_-}}  \right]  , \nonumber \\ \label{gammaimpar} 
\end{eqnarray}
\noindent where we have to recall the expression for the eigenvalues $\lambda_\pm$ in (\ref{eigenvalues}).

\section{Appendix: Covariance matrix coefficients } \label{V2MatrixCoeff}
In this appendix we show the explicit form of the coefficients of the covariance matrix ${\bf V}^{(2)}$.
\begin{eqnarray}
{V}^{(2)}_{11} &=& - \frac{\sinh ^2( \sqrt{ - \det{\bf a} } ) \left(a_{12}^2 l^4_1 + a_{22}^2 \hbar ^2\right)}{l^2_1 \det{\bf a} }+\frac{a_{12} l^2_1 \sinh (2 \sqrt{ - \det{\bf a} })}{\sqrt{ - \det{\bf a} }} + l^2_1 \cosh ^2(\sqrt{ - \det{\bf a} }) , \\
{V}^{(2)}_{22} &=& -\frac{\sinh ^2(\sqrt{ - \det{\bf c} }) \left(c_{12}^2 l^4_2 +c_{22}^2 \hbar ^2\right)}{l^2_2 \det{\bf c}}+\frac{c_{12} l^2_2 \sinh (2 \sqrt{ - \det{\bf c} })}{\sqrt{ - \det{\bf c} }}+ l^2_2 \cosh ^2(\sqrt{ - \det{\bf c} }) ,\\
{V}^{(2)}_{13} &=& \frac{\sinh (2 \sqrt{ - \det{\bf a} }) \left(a_{22} \hbar ^2-a_{11} l^4_1\right)}{2  l^2_1 \sqrt{ - \det{\bf a} } } + \frac{a_{12} \sinh ^2(\sqrt{ - \det{\bf a} }) \left(a_{11} l^4_1 + a_{22} \hbar ^2\right)}{l^2_1 \det{\bf a}} , \\
{V}^{(2)}_{24} &=& \frac{\sinh (2 \sqrt{ - \det{\bf c} }) \left(c_{22} \hbar ^2 - c_{11} l^4_2 \right)}{2  l^2_2 \sqrt{ - \det{\bf c} } } + \frac{c_{12} \sinh ^2(\sqrt{ - \det{\bf c} }) \left(c_{11} l^4_2 + c_{22} \hbar ^2\right)}{l^2_2 \det{\bf c}} , \\
{V}^{(2)}_{33} &=& - \frac{\sinh ^2(\sqrt{ - \det{\bf a} }) \left(a_{11}^2 l^4_1 + a_{12}^2 \hbar ^2\right)}{l^2_1 \det{\bf a}} + \frac{a_{12} \hbar ^2 \sinh (2 \sqrt{ - \det{\bf a} } )}{l^2_1 \sqrt{ - \det{\bf a}} } + \frac{\hbar ^2 \cosh ^2(\sqrt{ - \det{\bf a} })}{l^2_1}, \\
{V}^{(2)}_{44} &=& -\frac{\sinh ^2(\sqrt{ - \det{\bf c} }) \left(c_{11}^2 l^4_2 + c_{12}^2 \hbar ^2\right)}{l^2_2  \det{\bf c}} + \frac{c_{12} \hbar ^2 \sinh (2 \sqrt{ - \det{\bf c} } )}{l^2_2 \sqrt{- \det{\bf c}} } + \frac{\hbar ^2 \cosh ^2(\sqrt{ - \det{\bf c} })}{l^2_2}.
\end{eqnarray}

\end{document}